\documentclass[prl,onecolumn,superscriptaddress]{revtex4}
\usepackage{amsmath,amssymb,mathrsfs}

\usepackage{graphicx}
\usepackage{dcolumn}
\usepackage{bm}
\usepackage{paralist}
\usepackage[T1]{fontenc}
\usepackage[colorlinks,linkcolor=red,anchorcolor=blue,citecolor=blue]{hyperref}
\usepackage[caption=false,farskip=0pt,labelfont={bf}]{subfig}
\def\be{\begin{equation}}
\def\ee{\end{equation}}
\def\bea{\begin{eqnarray}}
\def\eea{\end{eqnarray}}

\begin{document}

\title{Excitation Spectra of one-dimensional  spin-1/2  Fermi gas with an attraction}

\author{Jia-Feng Pan}
\affiliation{State Key Laboratory of Magnetic Resonance and Atomic and Molecular Physics,
Wuhan Institute of Physics and Mathematics, Innovation Academy for Precision Measurement Science and Technology,  Chinese Academy of Sciences, Wuhan 430071, China}
\affiliation{Department of Fundamental and Theoretical Physics, Research School of Physics,
Australian National University, Canberra ACT 0200, Australia}

\author{Jia-Jia Luo}
\affiliation{State Key Laboratory of Magnetic Resonance and Atomic and Molecular Physics,
Wuhan Institute of Physics and Mathematics, Innovation Academy for Precision Measurement Science and Technology,  Chinese Academy of Sciences, Wuhan 430071, China}
\affiliation{University of Chinese Academy of Sciences, Beijing 100049, China}


\author{Xi-Wen Guan}
\email[]{xwe105@wipm.ac.cn}
\affiliation{State Key Laboratory of Magnetic Resonance and Atomic and Molecular Physics,
Wuhan Institute of Physics and Mathematics, Innovation Academy for Precision Measurement Science and Technology, Chinese Academy of Sciences, Wuhan 430071, China}
\affiliation{Department of Fundamental and Theoretical Physics, Research School of Physics,
Australian National University, Canberra ACT 0200, Australia}
\affiliation{NSFC-SPTP Peng Huanwu Center for Fundamental Theory, Xian 710127, China}

\date{\today}


\begin{abstract}
   Using exact  Bethe ansatz solution, we  rigorously study  excitation spectra of the spin-1/2 Fermi gas (called Yang-Gaudin model)  with an attractive interaction. 
   Elementary excitations of this model involve particle-hole excitation, hole excitation and adding particles in the Fermi seas of pairs and unpaired fermions. 
   The gapped magnon excitations in spin sector  show a ferromagnetic  coupling  to the Fermi sea of the single  fermions. 
    By numerically and analytically solving the Bethe ansatz equations  and the thermodynamic Bethe ansatz equations of this model, we obtain excitation energies for various polarizations in the phase of the Fulde-Ferrell-Larkin-Ovchinnikov (FFLO)-like  state. 
    For a small momentum (long-wavelength limit) and in the  strong interaction regime,  we analytically obtained their  linear dispersions with curvature corrections, effective masses as well as velocities  in particle-hole excitations of pairs and unpaired fermions.  
 Such a type of particle-hole excitations display a novel separation of collective motions  of bosonic modes within  paired and unpaired fermions.
Finally, we also discuss  magnon excitations in spin sector and the application of the Bragg spectroscopy for testing such separated charge excitation modes of  pairs and single fermions.

\end{abstract}

\maketitle
\section{Introduction}
The quantum many-body systems manifest abundant physical phenomena, such as Bose-Einstein Condensation (BEC), superfluidity, superconductivity, and quantum phase transition, etc.,  which are regarded as the emergent phenomena in modern physics. 
However, the complexity of  many-body systems, involving a huge internal degrees of freedom,  quantum statistics  and interaction,  alway brings us a formidable task  to access  their  physics of  interest. 
Among the known  many-body theories, there are two universal low energy effective  theories, i.e.  Fermi liquid  and Tomonaga-Luttinger liquid  that capture significant different  features of many-body correlations in   one dimension (1D)   and higher dimensions.
 Landau  Fermi Liquid theory \cite{baym2008landau,landau1957oscillations, landau1956fermi} remarkably describes metallic behaviour of  interacting fermions in 2D and 3D. 
In this theory  the concept of quasiparticles reveals the essence of  individual particle excitations close to the Fermi surface in  the interacting fermions. 
In contrast, when the degrees of freedom are  reduced to 1D, the elementary excitations in 1D systems strikingly  form  collective motions of bosons, which are  named  the Tomonaga-Luttinger liquid (TLL) in the long-wavelength limit \cite{tomonaga1950remarks,luttinger1963exactly}. 
In the TLL theory,  the low energy behaviour of 1D systems can be universally described by the bosonic fields of quantized sound waves or phonons. 
Early theoretical frame work of the TLL for various 1D  problems was developed by 
Lieb and Mattis\cite{mattis1994exact},  Haldane \cite{haldane1981luttinger} and   others, see reviews \cite{giamarchi2003quantum,Cazalilla2011one}. 
The TLL has become the main theme in the study of critical behaviour of 1D many-body systems. 
  

On the other hand, 
integrable models provide significant  insight into the emergent phenomena driven by interactions and quantum statistics  in 
low and higher dimensions \cite{essler2005one,Cazalilla2011one,guan2013fermi}. 
The quantum integrability should  trace back to H. Bethe's work in  1931 when he solved the eigenvalue problem 
of 1D spin-1/2 Heisenberg chain \cite{bethe1931theorie} by writing the wave function of the model  as the superposition of all possible  plane waves.
It was over 30 years after Bethe's work that this method was coined  as the BA (Bethe ansatz) by C. N. Yang and C. P. Yang in their series of publications in mid-60's \cite{YY-1,YY-2,YY-3}. 
Using Bethe's method,  Lieb and Liniger \cite{lieb1963exact} solved  the 1D Bose gas with a $\delta$-function interaction, which is now  called the Lieb-Liniger model.
In 1964 J. Mcguire\cite{mcguire1964study} studied  the 1D Fermi  gas with the $\delta$-function interaction  by considering the exact solution of  one spin-up fermion interacting with $N-1$  spin-down fermions. 
The exact solution of the 1D Fermi gas  with arbitrary numbers of  spin-up and spin-down particles  was  solved  by Yang \cite{yang1967some, yang1968s}, 
while at that time M. Gaudin \cite{gaudin1967systeme} obtained the BA solution for the model with a spin balance. 
This model is now named the Yang-Gaudin model. 
In C. N. Yang's seminal work \cite{yang1967some}, a key  discovery of the necessary condition for the BA solvability was surprisingly found. 
In 1972, R. J. Baxter \cite{baxter1972partition}  independently showed that such a factorization relation also occurred as the conditions for commuting transfer matrices in 2D vertex models in statistical mechanics.
It is now known as the Yang-Baxter equation, i.e. the factorization condition. 
The Yang-Baxter equation has laid out a profound legacy in a variety of fields in mathematics and physics. 
%

%

In 1969,  C. N. Yang and C. P. Yang \cite{yang1969thermodynamics}   obtained the thermodynamics of the Lieb-Liniger Bose gas  at  finite temperatures.
They found that  the thermodynamics of the model can be determined by  the minimisation conditions of the Gibbs free energy in terms of  the microscopic states determined by the BA equations.
M. Takahashi generalized  Yang and Yang's  method to deal with the thermodynamics of   the 1D Heisenberg spin chain and 1D Hubbard model though introducing string hypotheses \cite{takahashi2005thermodynamics, takahashi1970ground, takahashi1970many, takahashi1994one, takahashi1970magnetic}.
He coined the  method as the Yang-Yang thermodynamic Bethe ansatz (TBA), see a feature review 
\cite{guan2019professor}. 
Further developments of the TBA approach  have been made   in the study of  universal thermodynamics, Luttinger liquid, spin-charge separation, transport properties and critical phenomena for a wide range of low-dimensional quantum many-body systems, see  reviews \cite{guan2013fermi,Guan:2022}.


%
In the attractive regime, the Yang-Gaudin  model exhibits novel Fulde-Ferrell-Larkin-Ovchinnikov (FFLO) pairing correlation \cite{ff1964,lo1965}, i.e. coexistence of paired and unpaired fermions \cite{70-75in-Guan-Lin'spaper}.
Understanding of the FFLO  pairing behavior in 1D and higher dimensions is still  an open challenge  in condensed matter physics. 
The phase diagram of the attractive Fermi gas, consisting of a fully-paired state for an external field is less than the lower critical field $H_{c1}$, a fully-polarized state for  the magnetic field is greater than an upper critical field $H_{c2}$ and the FFLO-like state lies in between the two critical fields,  was predicted in  \cite{guan2007phase,orso2007attractive, hu2007phase} and experimentally confirmed  by R. Hulet's group in \cite{liao2010spin}.
 This novel phase diagram reveals striking features of thermodynamics, for instance, universal behaviour of  the  specific heat \cite{guan2011quantum}, the dimensionless ratios, such as Gr\"{u}neisen parameter \cite{peng2019gruneisen}  and Wilson ratios \cite{guan2013wilson,yu2016dimensionless}. 
The dark-soliton-like excitations in the Yang-Gaudin gas of attractively interacting fermions and the Lieb-Liniger gas \cite{shamailov2016dark,shamailov2019quantum} shed light on the nonlinear effects of many-body correlation.
Nevertheless, one expects that the elementary excitations in this attractive Yang-Gaudin model would  provide significant collective nature of multi-component TLLs with  pairing and depairing in thermodynamics and  dynamic response functions.   
This is  the major  research  of the following study in this paper. 

In Section \ref{sectionyanggaudin}, we will introduce the  BA  equations  and TBA equations, which will be used to accomplish our study of the excitation spectra of the Yang-Gaudin model with an attraction. 
In  Section \ref{sectionspectra}, we will present the particle-hole excitation spectra in paired and unpaired fermi seas. 
In Section \ref{sectionapp}, we will analytically derive dispersion relations with band curvature corrections, effective mass and sound velocities of pairs and unpaired fermions for the  Yang-Gaudin model with polarizations in  a strong coupling regime. 
In section \ref{sectionother}, we will discuss multiple particle-hole excitations and the magnon  excitations  in the FFLO-like phase.
The last Section is remained for our conclusion and discussion.

\section{Yang-Gaudin Model with an Attractive Interaction}
\label{sectionyanggaudin}
\subsection{Bethe Ansatz Equations and String Hypotheses}
The 1D two-component Fermi gas with a delta-function interaction  is called the Yang-Gaudin 
model\cite{yang1967some, yang1968s, gaudin1967systeme}.
Its Hamiltonian is given by 
\begin{equation}
    \mathcal{H}=-\sum_{i=1}^{N} \frac{\partial^{2}}{\partial x_{i}{ }^{2}}+4 c 
    \sum_{i<j} \delta\left(x_{i}-x_{j}\right)-\mu_{0} H\left( N-2N_\downarrow\right),
\end{equation}
where $N$ is the number of fermions, $N_\downarrow$ is the number of down-spin fermions,
$H$ is the external magnetic field, $c<0$ for attractive interaction,  
and herewith we set $\hbar^2=2m=1$. We always choose an upward magnetic
field so that spin-down fermions are less than spin-up ones. As a result, each spin-down
fermion can be paired with a spin-up fermion and form a bounded state.
While  the remaining spin-up fermions are unpaired and in polarized state.
Denote the number of paired fermions as $M$, then that of unpaired fermions is $N-2M$.

The quasimomenta $\{k_j\}$ of the fermions and the rapidities $\{\Lambda_{\alpha}\}$ 
of the spin-down fermions are given by the BA Equations
\begin{equation}
    \begin{aligned}
        &e^{i k_{j} L}=\prod_{\alpha=1}^{N_\downarrow} 
        e\left(\frac{k_{j}-\Lambda_{\downarrow\alpha}}{c}\right),  \\
        &\prod_{j=1}^{N} e\left(\frac{\Lambda_{\downarrow\alpha}-k_{j}}{c}\right)=
        \prod_{\beta \neq \alpha} e\left(\frac{\Lambda_{\downarrow\alpha}-\Lambda_{\downarrow\beta}}{2 c}\right) 
        \label{ba}
    \end{aligned}
\end{equation}
for $j=1,2,\cdots,N$ and $\alpha=1,2,\cdots,N_\downarrow$, where 
$e\left(x\right)=\exp\left(i\pi-2i\arctan{x}\right)$.

Takahashi\cite{takahashi2005thermodynamics, takahashi1970ground, takahashi1970many, 
takahashi1994one, takahashi1970magnetic} introduced the following spin string hypotheses for the root patterns of the BA equations (\ref{ba}):\\
1. Complex $\Lambda_{\downarrow\alpha}$ always forms a bound state with several other $\Lambda_{\downarrow\alpha}$'s. 
In this set of $n$-$\Lambda_{\downarrow\alpha}$'s the real parts are the same 
and the imaginary parts are $(n-1) c i,(n-3) c i, \cdots,-(n-1) c i$ for each of them
within the accuracy of $\mathcal{O}(\exp (-\delta N))$, 
where $\delta$ is a positive number.
This bound state of $n$-$\Lambda_{\downarrow\alpha}$'s forms an $n$-string.
In the following discussion, we will  denote each of these complex $n$-$\Lambda_{\downarrow\alpha}$ strings 
as $\Lambda_\alpha{}^{n,j}$, 
where $j=1,2,,\cdots,n$. 
We suppose that  there are $M_n$-$n$-strings, 
and denote their  real part of each $n$-string as $\Lambda_\alpha{}^n$, 
where $\alpha=1,2,\cdots,M_n$. 
Than we  can express  the $n$-string as 
\begin{equation}
    \Lambda_{\alpha}^{n, j}=
    \Lambda_{\alpha}{ }^{n}+(n+1-2 j) c i+\mathcal{O}(\exp (-\delta N)), j=1,2, \cdots, n.
\end{equation}
2. For an attractive interaction $c<0$,  a pair of quasi-momenta  form a charge bound sate, namely 
$k_{\alpha}$ and  its complex conjugate $\bar{k}_{\alpha}$ have a common  real part $\Lambda$,  i.e.,
\begin{equation}
\begin{aligned}
    &k_\alpha=\Lambda_\alpha+i\left|c\right|
    +\mathcal{O}\left(\exp \left(-\delta L\right)\right),\\
    &\bar{k}_{\alpha}=\Lambda_\alpha-i\left|c\right|
    +\mathcal{O}\left(\exp \left(-\delta L\right)\right).
\end{aligned}
\end{equation}

Supposing that  there are $M$ charge bound pairs and  $\left(N-2M\right)$ single fermions with 
the quasi-momenta $\left\{ k_j\right\}$ with  $j=1,2,\cdots,N-2M$.
Then  the BA equations (\ref{ba}) can be rewritten in the following form \cite{takahashi1994one} 
\begin{equation}
    \begin{aligned}
        &2 \Lambda_{\alpha} L=2 \pi J_{\alpha}+
        \sum_{j=1}^{N-2 M} 
        \theta\left(\frac{\Lambda_{\alpha}-k_{j}}{|c|}\right)+
        \sum_{\beta=1}^{M} 
        \theta\left(\frac{\Lambda_{\alpha}-\Lambda_{\beta}}{2|c|}\right), 
        \quad \alpha=1,2, \cdots, M \\
        &k_{j} L=2 \pi I_{j}+
        \sum_{\alpha=1}^{M} 
        \theta\left(\frac{k_{j}-\Lambda_{\alpha}}{|c|}\right)+
        \sum_{n=1}^{\infty} \sum_{\alpha=1}^{M_{n}} 
        \theta\left(\frac{k_{j}-\Lambda_{\alpha}{ }^{n}}{n|c|}\right), 
        \quad j=1, \cdots, N-2 M \\
        &\sum_{j=1}^{N-2 M} 
        \theta\left(\frac{\Lambda_{\alpha}{ }^{n}-k_{j}}{n|c|}\right)=
        2 \pi J_{\alpha}{ }^{n}+
        \sum_{m=1}^{\infty} \sum_{\beta=1}^{M_{m}} 
        \Theta_{n m}\left(\frac{\Lambda_{\alpha}{ }^{n}-\Lambda_{\beta}^{m}}{|c|}\right), 
        \quad 
        \begin{array}{l}
        \alpha=1,2, \cdots, M_{n} \\
        n=1,2, \cdots.
        \end{array}
    \end{aligned}
    \label{eqlogBA}
\end{equation}
where $\theta\left(x\right) = 2 \arctan x,-\pi<\theta < \pi$, and
\begin{equation}
\begin{aligned}
&\Theta_{n m}\left(x\right) \\=&
\begin{cases}
    \theta\left(\frac{x}{|n-m|}\right)+2 \theta\left(\frac{x}{|n-m|+2}\right)
    +2 \theta\left(\frac{x}{|n-m|+4}\right) +\cdots+2 \theta\left(\frac{x}{n+m-2}\right)
    +\theta\left(\frac{x}{n+m}\right) &\text { for } n \neq m \\
    2 \theta\left(\frac{x}{2}\right)+2 \theta\left(\frac{x}{4}\right)+\cdots
    +2 \theta\left(\frac{x}{2 n-2}\right)+\theta\left(\frac{x}{2 n}\right) 
    & \text { for } n=m
\end{cases}
.
\end{aligned}
\end{equation}
Here $J_{\alpha}$ is integer (half-odd integer) for $N-M$ odd (even), 
$I_{j}$ is integer (half-odd integer) for $M+M_{1}+M_{2}+\cdots$ even (odd), and 
$J_{\alpha}{ }^{n}$ is integer (half-odd integer) for $N-M_{n}$ odd (even). 
$J_{\alpha}{ }^{n}$ should satisfy the condition
\begin{equation}
    \left|J_{\alpha}{ }^{n}\right| \leqq \frac{1}{2}\left(N-2 M-
    \sum_{m=1}^{\infty} t_{n m} M_{m}\right),\quad
    t_{mn}=2\mathrm{min}\left(n,m\right)-\delta_{n,m}.
    \label{J^n}
\end{equation}
Eq.(\ref{eqlogBA}) give quantum numbers $J_{\alpha}$, $I_j$, and 
$J_{\alpha}{ }^{n}$ for charge sector of paired fermions, charge sector of unpaired 
fermions, and spin sector of unpaired fermions, respectively.
It turns out that for  each pair $k_{\alpha}$ and $\bar{k}_{\alpha}$  are given by  $\Lambda_{\alpha}$, 
representing quasi-momenta of pairs. Whereas $k_j$ represent 
quasi-momenta of $N-2M$ unpaired fermions and $\Lambda_{\alpha}{ }^{n}$ represent $M_n$ spin wave bound states  (the length-$n$ strings).

In the  thermodynamic limit, i.e., $L,\, N\rightarrow \infty$, these quantum numbers could be treated as functions of  continuous variables $k$s, which satisfy
\begin{equation}
\begin{aligned}
    &\mathrm{d} J_{\alpha}\left(k\right)
    =L \left(\sigma\left(k\right)+\sigma^{h}\left(k\right)\right) 
    \mathrm{d} k,\\
    &\mathrm{d} I_{j}\left(k\right)
    =L \left(\rho\left(k\right)+\rho^{h}\left(k\right)\right) \mathrm{d} k,\\
    &\mathrm{d} J_{\alpha}{ }^{n}\left(k\right)
    =L \left(\sigma_{n}\left(k\right)+\sigma_{n}^{h}\left(k\right)\right) \mathrm{d} k,
\end{aligned}
\label{eqdefinedistribution}
\end{equation} 
where $\sigma,\rho,\sigma_n$ denote distribution functions of  bound pairs, single particles and length-$n$ spin strings, respectively. 
While 
$\sigma^{h}, \rho^{h}, \sigma_{n}^{h}$ denote the distribution functions of their corresponding holes.
The continuous $k$ stands for discrete quasi-momenta $k_j$, $\Lambda_{\alpha}$ and $\Lambda_{\alpha}{ }^{n}$. 
It follows that the BA equations  (\ref{ba}) become as  \cite{takahashi1994one}
\begin{equation}
    \begin{aligned}
        &\frac{1}{\pi}=\sigma+\sigma^{n}+[2] \sigma+[1] \rho \\
        &\frac{1}{2 \pi}=\rho+\rho^{h}+[1] \sigma+\sum_{n}[n] \sigma_{n} \\
        &{[n] \rho=\sigma_{n}{ }^{h}+\sum_{m} A_{n m} \sigma_{m}},
        \label{6.7}
    \end{aligned}
\end{equation}
where $[n]$ is an operator defined by
\begin{equation}
\begin{aligned}
    &[n] f\left(k\right) =  \int_{-\infty}^{\infty} a_n\left(k-k'\right) 
    f\left(k^{\prime}\right) d k^{\prime}, \quad
    a_n\left(k\right)=\frac{1}{\pi}\frac{n|c|}{\left(n c\right)^{2}+k^{2}}, \\
    &[0] f\left(k\right) = f\left(k\right).
\end{aligned}
\end{equation}
Here 
\begin{equation}
    A_{n m} =[|n-m|]+2[|n-m|+2]+2[|n-m|+4]+\cdots+2[n+m-2]+[n+m].
\end{equation}


\subsection{Thermodynamical Bethe Ansatz Equations and Thermodynamic Quantities}
Building on microscopic state energies, we  may further define dressed energies of  pairs $\varepsilon=T\ln{\eta}$,  single fermions 
$\kappa=T\ln{\zeta}$, and length-$n$ strings  $\varepsilon_n=T\ln{\zeta_n}$, respectively. 
Here we denoted the quantities $\zeta=\rho^{h} / \rho,$ $\eta^=\sigma^{h} / \sigma,$ 
and $\eta_{n}=\sigma_{n}^{h} / \sigma_{n}$. 
Minimizing  thermodynamic potential $\Omega \equiv E-T S-A N-H S_Z$ with respect to the densities though Eq.(\ref{6.7}), 
i.e., $\delta\Omega=0$,   we may obtain the following \textcolor{blue}{TBA} Equations \cite{takahashi1994one}
\begin{equation}
    \begin{aligned}
        &\ln \eta^=\frac{2\left(k^{2}-A-c^{2}\right)}{T}
        +[2] \ln \left(1+\eta^{-1}\right)+[1] \ln \left(1+\zeta^{-1}\right), \\
        &\ln \zeta=\frac{k^{2}-A-\mu_{0} H}{T}
        +[1] \ln \left(1+\eta^{-1}\right)
        -\sum_{n=1}^{\infty}[n] \ln \left(1+\eta_{n}^{-1}\right), \\
        &\ln \left(1+\eta_{n}\right)=\frac{2 n \mu_{0} H}{T}
        +[n] \ln \left(1+\zeta^{-1}\right)
        +\sum_{m=1}^{\infty} A_{n m} \ln \left(1+\eta_{m}^{-1}\right),
        \label{6.10}
    \end{aligned}
\end{equation}
where $n=1,\dots, \infty$. 
In the above equations, $T$, $A$ and $H$ stand for temperature, chemical potential and magnetic field, respectively. 
Accordingly, we can give the equation of state, namely the pressure $p$ is given by 
\begin{equation}
p=-\Omega / L=T \int_{-\infty} ^{\infty}  \ln \left(1+\eta^{-1}\right) \frac{\mathrm{d}k}{\pi}
+T \int _{-\infty} ^{\infty}  \ln \left(1+\zeta^{-1}\right) \frac{\mathrm{d}k}{2 \pi}, 
\label{equationofstate}
\end{equation}
The pressure of this system is simply the sum of two terms, the term regarding the paired Fermi sea  and another regarding the unpaired sea, respectively.
Other thermodynamic quantities can be calculated through the usual thermodynamics  relations: 
\begin{equation}
    N/L=\frac{\partial p}{\partial \mu}, \quad 
    S_z/L=\frac{\partial p}{\partial H}, \quad 
    S/L=\frac{\partial p}{\partial T}, \quad 
    \kappa=\frac{\partial^2 p}{\partial A^2}, \quad 
    \chi=\frac{\partial^2 p}{\partial H^2}, \quad 
    c_{V}=T \frac{\partial^2 p}{\partial T^2},
\end{equation}
where the last three quantities are the compressibility, magnetic susceptibility, and specific 
heat.

At the zero temperature $T\rightarrow0$, due to the ferromagnetic ordering,  we observe that  $\eta_n\rightarrow\infty$ and $\sigma_n=0$
for $n=1,2,\cdots$ and  the TBA equations (\ref{6.10}) reduce to 
\begin{equation}
    \begin{aligned}
        &\varepsilon\left(k\right)=2\left(k^{2}-A-c^{2}\right)
        -[2] \varepsilon^{-}\left(k\right)-[1] \kappa^{-}\left(k\right), \\
        &\kappa\left(k\right)=k^{2}-A-\mu_{0} H-[1] \varepsilon^{-}\left(k\right),
    \end{aligned}
    \label{eqdressed1}
\end{equation}
where the superscript '$-$' means the corresponding quantities are token the negative parts. 
Since $\varepsilon_n{}^-\rightarrow 0$ for $n\ge 1$  in the limit $T\rightarrow 0$, these equations are referred to the zero temperature dressed energy equations. 
In this case, the Fermi points  $B$, $Q$ are determined by $\varepsilon\left(B\right)=0$ and $\kappa\left(Q\right)=0$.
It turns out that $\varepsilon\left(k\right)$ and $\kappa\left(k\right)$ are monotonically increasing functions of $k^2$.
Therefore, $B$ and $Q$ are
Fermi surfaces referring to the continuous quasimomentum of paired and unpaired fermions. 
 They can also  be determined by the relations
\begin{eqnarray}
n&\equiv:& \frac{N}{L}= 2\int_{-B}^{B}\sigma(k)dk+\int^{Q}_{-Q}\rho(k)dk,\nonumber \\
n_{\downarrow}&\equiv: & \frac{N_{\downarrow}}{L}=\int_{-B}^{B}\sigma(k)dk.\label{density-a}
\end{eqnarray} 
Thus the distribution functions are given by 
\begin{equation}
    \begin{aligned}
        &\sigma\left(k\right)+\sigma^{h}\left(k\right)=\frac{1}{\pi} 
        - \int_{-B}^{B} a_2\left(k-k'\right)\sigma\left(k'\right)\mathrm{d}k'
        - \int_{-Q}^{Q} a_1\left(k-k'\right)\rho\left(k'\right)\mathrm{d}k', \\
        &\rho\left(k\right)+\rho^h \left(k\right)=\frac{1}{2 \pi}
        - \int_{-B}^{B} a_1\left(k-k'\right)\sigma\left(k'\right)\mathrm{d}k',
    \end{aligned}
    \label{eqdistribution}
\end{equation}
where $\sigma_n=0$ for the ground state and spin sector is completely gapped. 

Moreover, at zero temperature limit $T\rightarrow 0$, there exists the phase transition among fully paired phase, fully-polarized phase, partially-polarized phase, and vacuum phase, which will be further studied  in Section \ref{sectorstrongcoupling}.  
The fully paired phase can be regarded as the Bardeen-Cooper-Schrieffer(BCS) like phase, which 
is expected to manifest the first-type superconductivity. 
The phase diagram are  given in \cite{orso2007attractive,hu2007phase,guan2011quantum, yin2011quantum}.
Fulde and Ferrell\cite{ff1964} and 
Larkin and Ovchinnikov\cite{lo1965} predicted the  exotic superfluid phase, which is now called the FFLO phase. 
The partially polarized phase in the attractive Yang-Gaudin model  is composed of BCS-like pairs and unpaired fermions, presenting 
 a 1D-analogy of the FFLO phase.
The  FFLO-like phase diagram  of  the attractive Fermi gas was experimentally confirmed by Hulet's group  \cite{pini2021strong}.

\section{Spectra of One-particle-hole Excitations}
\label{sectionspectra}
\subsection{Ground State}
\label{sectiongroundstate}
Consider the partially polarized phase, where paired and unpaired fermions coexist.
The total momentum of the system is given by 
\begin{equation}
    K=2\sum_{\alpha=1}^{M}\Lambda_\alpha+\sum_{j=1}^{N-2M}k_j
    =\frac{2\pi}{L} \left(\sum_{\alpha=1}^{M}J_\alpha 
    +\sum_{j=1}^{N-2M}I_j +\sum_{n=1}^{\infty}\sum_{\alpha=1}^{M_n}J_{\alpha}{}^{n}\right)
    =K_{b}+K_{u}+K_{s,u}.
    \label{totalmomentum}
\end{equation}
Due to the ferromagnetic ordering in spin sector, the distribution functions $\sigma_n=0$  with $n=1,2,\cdots$ at the ground state.  
In other words, there is no particle for the length $n$-strings, i.e., $M_n$=0. 
Therefore, the spin sector of unpaired fermions does not have  contribution to the total momentum, i.e., $K_{s,u}=0$ at the ground state.

\begin{figure}[ht] 
    \centering
    \includegraphics[width=0.7\textwidth]{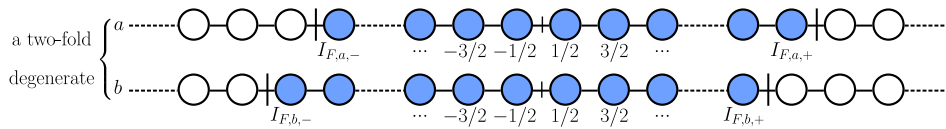}
    \caption[schematic illustration  of the two-fold degeneration.]
    {\textbf{Schematic illustration  of the two-fold degeneration}.  
    $I_{F,a,-}$ and $I_{F,a,+}$, (or $I_{F,b,-}'$ and $I_{F,b,+}'$ ) give respectively  the Fermi points of unpaired (paired)  sector.
    In this case, the particle number  is  odd while the corresponding  quantum number $I_j$s take  half-odd integers.
    It appears that these quantum numbers cannot distribute symmetrically.  
    Therefore, the total momentum is non-zero and  the ground state is two-fold degenerate.}
    \label{degenerate}
    \end{figure}
    
    \begin{figure}[ht] 
    \centering
    \subfloat[density distributions]{
        \includegraphics[width=0.48\textwidth]{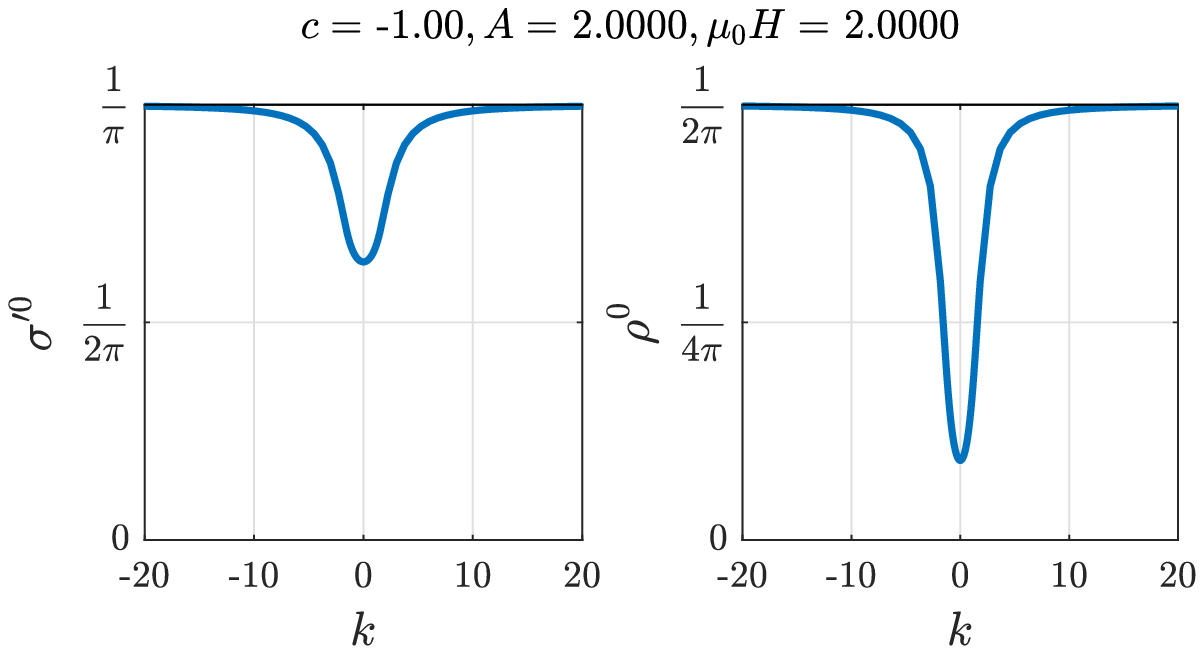}
    }
    \subfloat[dressed energies]{
        \includegraphics[width=0.48\textwidth]{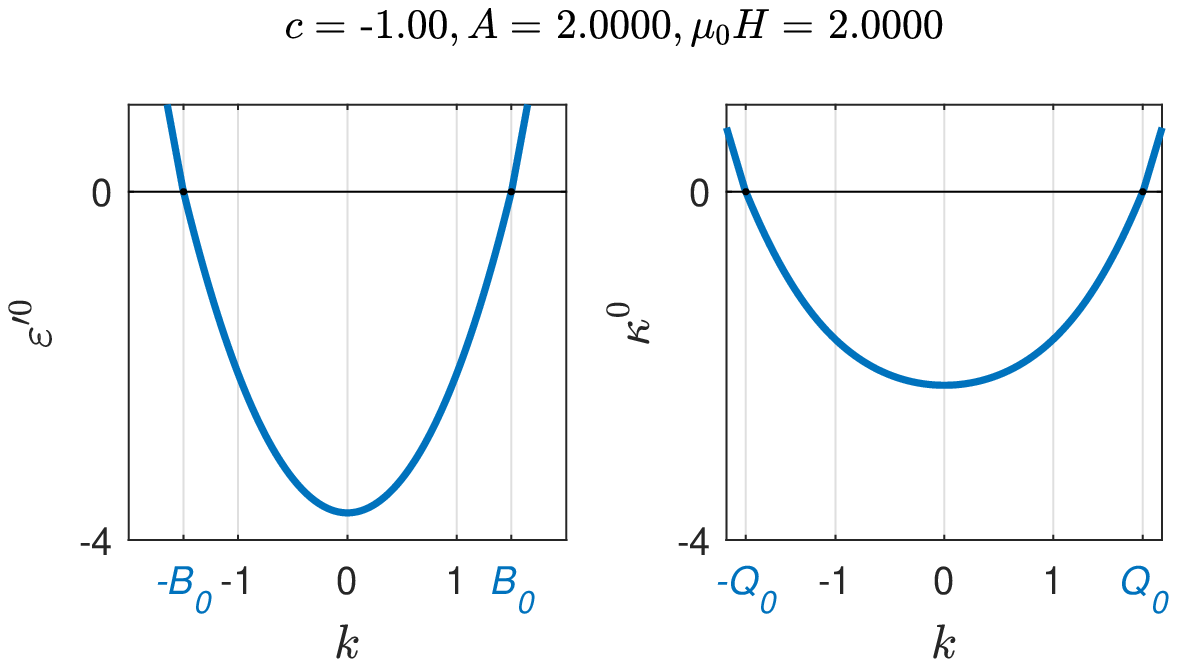}
    }
    \caption[ Density distribution functions and dressed energies]
    {\textbf{(a) Density distribution functions and (b) Dressed energies at zero temperature.}
    (a) For $k\rightarrow\pm\infty$, the density distributions of paired  fermions $\sigma^0$ and unpaired fermions $\rho^0$ tend  to $1/\pi$ or $1/(2\pi)$, respectively. 
    (b) The dressed energies of paired and unpaired fermions  are  monotonically increasing with $|k|$.
    Therefore, the Fermi points referring to the continuous quasi-momenta can be well defined  by the conditions  $\varepsilon^0(\pm B_0)=0$ and $\kappa^0(\pm Q_0)=0$, respectively.}
    \label{distribution}
\end{figure}

For  the ground state, both paired and unpaired fermions should occupy those lowest-energy
states, i.e., states with the lowest absolute quantum numbers (quasi-momenta). 
Consequently, most contributions of particles to total momentum are neutralized in the sum of positive and negative quantum numbers.
The total momentum is not zero only if the distribution of quasi-momenta is not symmetric
around the zero quasi-momentum. 
Fig.\ref{degenerate} presents an example of non-zero total momentum, which essentially shows  a two-fold degeneration.

There are totally $M$ paired fermions, 
whereas $J_\alpha$s  take  integers (half-odd integers) for $N-M$ odd (even). 
Therefore, $J_\alpha$s are  symmetrically distributed if and only if $N$ is even.
Similarly, there are $N-2M$ unpaired fermions, 
whereas $I_{j}$s  take integers (half-odd integers) for $M+\sum M_n=M$ are  even (odd).
Therefore, $I_{j}$ is symmetrically distributed if and only if $N+M$ is odd, see Eq.(\ref{eqlogBA}).
Consequently, 
\begin{equation}
    K_{b}=
    \begin{cases}
        0 &\text{for }N\text{ even},\\
        \pm \pi n_b &\text{for $N$ odd},
    \end{cases}
    \qquad
    K_{u}=
    \begin{cases}
        0            &\text{for $N+M$ odd},\\
        \pm \pi n_u &\text{for $N+M$ even}, 
    \end{cases}
    \label{eqsconfiguration}
\end{equation}
where $n_b=M/L$ and $n_u=(N-2M)/L$ are particle densities of paired and unpaired sectors, respectively. 

Introduce the superscript "$0$" to denote the ground state. 
Accordingly, the dressed-energies Eq.(\ref{eqdressed1}) and the density distribution functions and Eq.(\ref{eqdistribution}) are respectively written as
\begin{equation}
    \begin{aligned}
        \sigma^{0}\left(k\right)&=\frac{1}{\pi} 
        - \int_{-B_0}^{B_0} a_2\left(k-k'\right)\sigma^{0}\left(k'\right)\mathrm{d}k'
        - \int_{-Q_0}^{Q_0} a_1\left(k-k'\right)\rho^{0}\left(k'\right)\mathrm{d}k' \\
        \rho^0\left(k\right)&=\frac{1}{2 \pi} 
        - \int_{-B_0}^{B_0} a_1\left(k-k'\right)\sigma^{0}\left(k'\right)\mathrm{d}k'
    \end{aligned}
    \label{3.distrib}
\end{equation}
and
\begin{equation}
    \begin{aligned}
        \varepsilon^{0}\left(k\right)&=2\left(k^{2}-A-c^{2}\right)
        - \int_{-B_0}^{B_0} a_2\left(k-k'\right)\varepsilon^{0}\left(k'\right)\mathrm{d}k'
        - \int_{-Q_0}^{Q_0} a_1\left(k-k'\right)\kappa^{0}\left(k'\right)\mathrm{d}k' \\
        \kappa^0\left(k\right)&=k^{2}-A-\mu_{0} H-\int_{-B_0}^{B_0} a_1\left(k-k'\right)
        \varepsilon^{0}\left(k'\right)\mathrm{d}k',
    \end{aligned}
    \label{3.dressed}
\end{equation}
both of which are plotted in  Fig. \ref{distribution} with certain values of the interaction strength, chemical potential and magnetic field, i.e. $\{c,A,\mu_0H\}$. 

In terms of the additivity of the total momentum in Eq.(\ref{totalmomentum}), we can naturally define the Fermi points, $k_{F,b,\pm}$ and $k_{F,u,\pm}$, referring to the discrete quasimomenta of both paired and unpaired sectors, by maximal and minimal quantum numbers within occupied $\{J_\alpha\}$ and $\{I_j\}$, respectively.
For $J_\alpha$ symmetrically distributed, $\left|k_{F,b,\pm}\right|=2\pi\left|J_{F,\pm}\right|/L=\pi(M-1)/L\approx\pi n_b$ for a thermodynamical limit, i.e. $N$, $L\rightarrow\infty$; whereas for the asymmetrical case, we still have $\left|k_{F,b,+}\right|=\pi(M-1\pm1)/L\approx\pi n_b$ and $\left|k_{F,b,-}\right|=\pi(M-1\mp1)/L\approx\pi n_b$, where the additional terms $\pm\pi/L$ emerging due to the two-fold degeneration are negligible for the thermodynamical limit.
With similar regard, we have $k_{F,u,\pm}=2\pi I_{F,\pm}\approx\pi n_u$ for the Fermi points referring to the unpaired fermions.
The Fermi points are referred  to the conditions $\varepsilon^{0}\left(Q_0\right)=0$ and $  \sigma^{0}\left(B_0\right)=0$, where $B_0$ and $Q_0$ are different from the   quasimomenta $k_{F,b,\pm}$ and $k_{F,u,\pm}$ of pairs and unpaired fermions. 

\subsection{Sound Velocity and Effective Mass for One-Particle-hole Excitation}
\label{sectionvm}
\label{sectionphex}
\begin{figure}[ht] 
    \centering
    \includegraphics[width=0.65\textwidth]{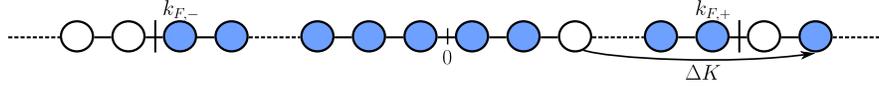}
    \caption[Schematic illustration of the one-particle-hole excitation.  ]
    {\textbf{Schematic illustration of the one-particle-hole excitation:} Moving one particle within the two Fermi points to outside  the Fermi sea. 
    }
    \label{figphex}
\end{figure}

\begin{figure}[htbp]
    \centering
    \includegraphics[width=0.48\textwidth]{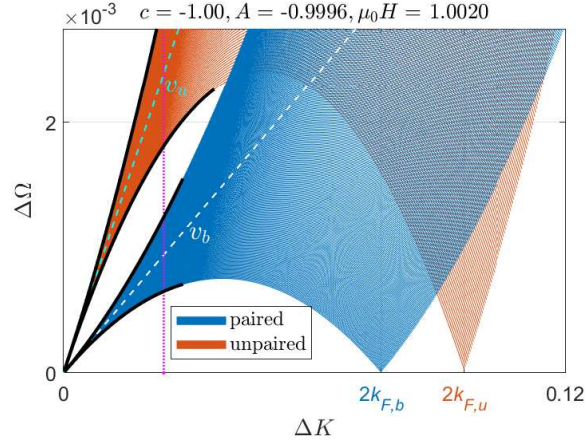}
    \caption[One-particle-hole excitation spectra of paired  and unpaired fermions. ]
    {\textbf{One-particle-hole excitation spectra of paired (blue) and unpaired  (dark brown) fermions.}
    $k_{F,b}$ ($v_b$) and $k_{F,u}$ ($v_u$)  are the Fermi momenta (velocities) of paired and unpaired fermions, respectively. 
    For small $\Delta K$,  the  paired and unpaired excitation spectra tend to the linear dispersions with different velocities    $v_b$ and $v_u$,  which are indicated by the white and green dashed lines, respectively.  The curvature effects of these dispersion can be conceived from the effective masses which were further studied in Section \ref{sectionapp}.  In this figure, we set $c=-1.00$, $A=-0.9996$, $\mu_0H=1.0020$ for our numerical calculation with the dispersions represented by Eqs.(\ref{eqexcitedenergyunpired}), (\ref{excitedmomentumunpaired}) and (\ref{eqdispersionpaired}). 
    }
    \label{1phboundary}
\end{figure}
    
\begin{figure}[htbp]
    \centering
    \subfloat[$k_{F,b}= k_{F,u}$]{
        \includegraphics[width=0.48\textwidth]{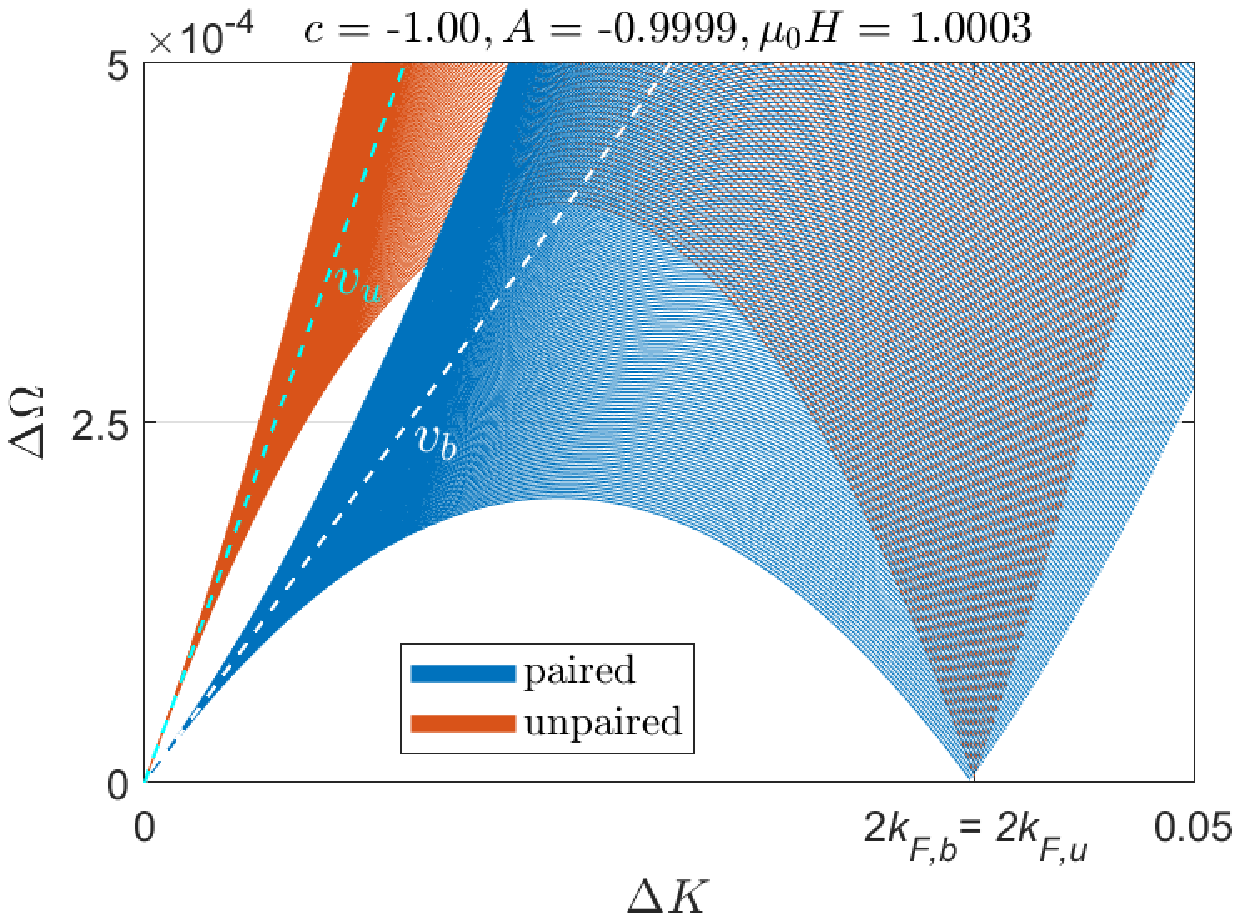}
    }
    \subfloat[$v_{u}= v_{b}$]{
        \includegraphics[width=0.48\textwidth]{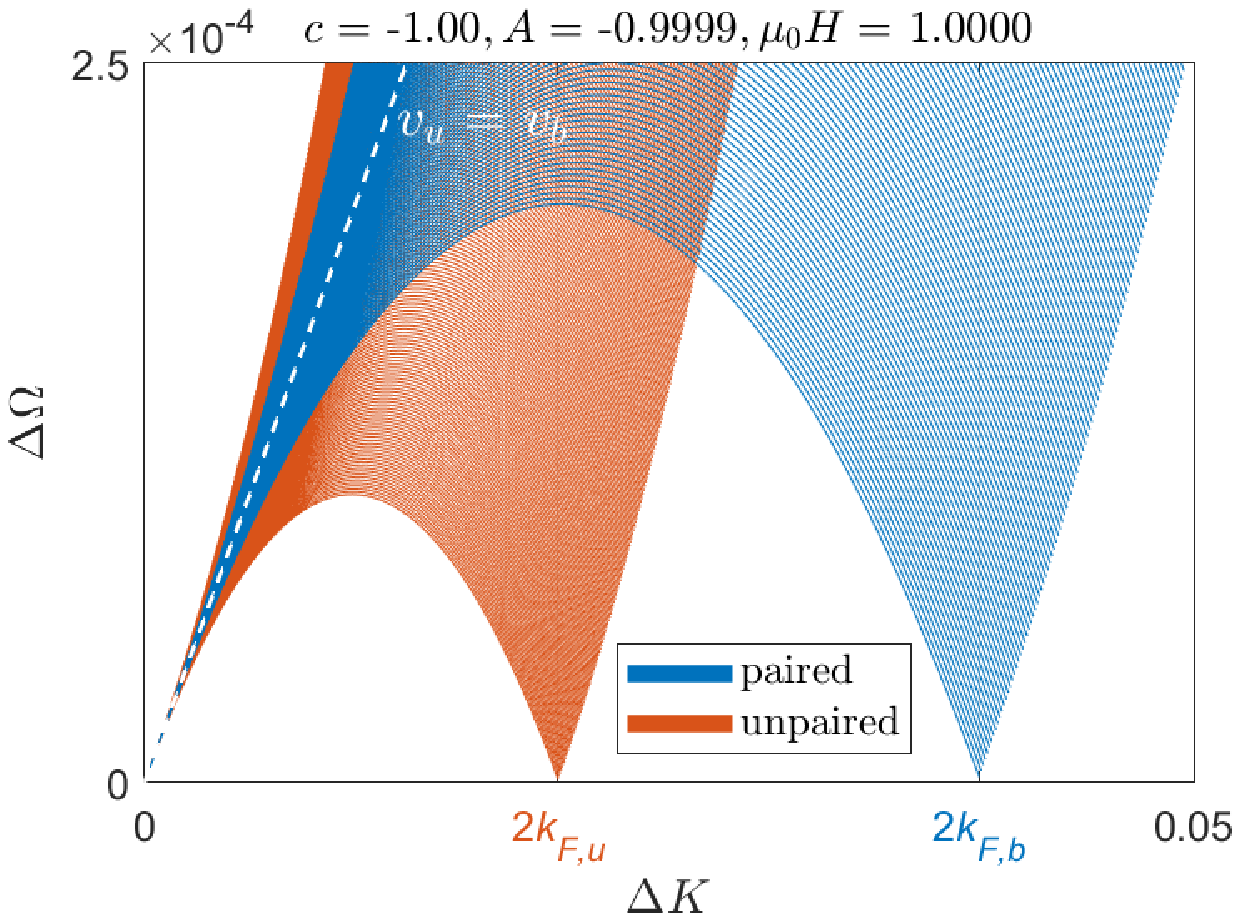}
    }
    \caption[One particle-hole excitation spectra of paired (blue) and unpaired (dark brown) fermions ]
    {\textbf{One-particle-hole excitation spectra of paired (blue) and unpaired (dark brown) fermions  with $k_{F,b} = k_{F,u}$ (a)  and $v_{u}=2v_{b}$ (b)}.  In (a), we set $c=-1.00$, $A=-0.9999$, $\mu_0H=1.0003$ for our numerical calculation with the Eqs.(\ref{eqexcitedenergyunpired}), (\ref{excitedmomentumunpaired}) and (\ref{eqdispersionpaired}); In (b), we set $c=-1.00$, $A=-0.9999$, $\mu_0H=1.0003$.
    These parameters are artificially selected, which we will explain in Section \ref{sectionapp}.
    }
    \label{1ph}
\end{figure}

Consider one particle-hole excitation of unpaired fermions, where one quasiparticle with a quasimomentum $k_h$ inside the Fermi points is excited outside the Fermi points with a quasimomentum $k_p$, i.e.,
$$
\rho\left(k\right)\neq0, 
\text{ only for } k\in\left[Q_-,k_h\right)\cup\left(k_h,Q_+\right] \text{ and } k= k_p.
$$
Define $\bar{\rho}\left(k\right)=\rho\left(k\right)+\rho^h\left(k\right)$, which is changed by $\Delta\bar{\rho}\left(k\right)$ after the excitation. 
Since there is exactly one excited hole inside the Fermi points and one quasiparticle outside, 
$$
\int_{-Q}^{Q}\Delta\bar{\rho}\left(k\right)L\mathrm{d}k=-1, \quad
\left(\int_{-\infty}^{-Q}+\int_{Q}^{\infty}\right)\Delta\bar{\rho}\left(k\right)L\mathrm{d}k=1.
$$
Therefore, $\Delta\bar{\rho}\left(k\right)$ is the sum of two $\delta$-functions, $-\delta(k-k_h)/L$ and $\delta(k-k_p)/L$. 
Also define $\bar{\sigma}\left(k\right)=\sigma\left(k\right)+\sigma^h\left(k\right)$.
Consequently, the distribution functions Eq.(\ref{eqdistribution}) are rewritten as
\begin{equation}
    \begin{aligned}
        \bar{\sigma}\left(\lambda\right)=&\frac{1}{\pi} 
        - \int_{B_-}^{B_+} a_2\left(\lambda-k'\right)\sigma\left(k'\right)\mathrm{d}k'
        - \int_{Q_-}^{Q_+} a_1\left(\lambda-k'\right)\bar{\rho}\left(k'\right)\mathrm{d}k'
        +\frac{1}{L}a_1\left(\lambda-k_h\right)-\frac{1}{L}a_1\left(\lambda-k_p\right) \\
        \bar{\rho}\left(k\right)=&\frac{1}{2 \pi} 
        - \int_{B_-}^{B_+} a_1\left(k-k'\right)\sigma\left(k'\right)\mathrm{d}k',
        \label{distributionintroducedelta}
    \end{aligned}
\end{equation}
where $B_\pm$ and $Q_\pm$ are fermi surfaces of the excited system. 
In TBA approach, the equilibrium state of the system is determined 
by minimize free energy $\Omega=E-\mu N-T S$, so, the excited free energy is
\begin{equation}
    \begin{aligned}
        \frac{\Delta \Omega}{L} =\frac{E}{L}-\frac{E_{G}}{L}
        =&\int_{Q_-}^{Q_+}\rho\left(k\right) 
        \left(k^{2}-\mu_0 H\right) \mathrm{d} k
        -\int_{-Q_0}^{Q_0} \rho^{0}\left(k\right) \left(k^{2}-\mu_0 H\right) \mathrm{d} k \\
        &+2\int_{B_-}^{B_+} \sigma\left(\lambda\right) \left(\lambda^2-c^2\right) \mathrm{d}\lambda 
        -2\int_{-B_0}^{B_0}\sigma\left(\lambda\right)\left(\lambda^2-c^2\right) \mathrm{d}\lambda 
    \end{aligned}
    \label{excitedenergytoberewritten}
\end{equation}
For the Yang-Gaudin model with repulsive interactions, it is proved \cite{he2020emergence} that the excited energy can be expressed in the dressed energies. 
Using  a similar method and after a lengthy calculation,   we may obtain excitation spectra in the attractive Fermi gas  (see Appendix.\ref{Express the excited energy in dressed energies}), namely 
\begin{equation}
    \begin{aligned}
        \Delta \Omega
        =\kappa^0\left(k_p\right)-\kappa^0\left(k_h\right).
    \end{aligned}
    \label{eqexcitedenergyunpired}
\end{equation}
Since the particle numbers $N,\,M,$ and $M_n$ are conserved, the configurations  of quasimomenta remain  unchanged. 
In other words, if the  positions of quasi-momenta  are originally symmetrically (asymmetrically) distributed with respect to  the zero-point, they are still symmetrically (asymmetrically) distributed after an excitation. 
Therefore, the excited momentum of the systems is  equal to  the quasi-momentum of the excited particle, namely,
\begin{equation}
    \Delta K
    =\frac{2\pi}{L}\left(I_p-I_h\right)
    =2\pi\left[\int_{0}^{k_p} \rho^0\left(k\right) \mathrm{d}k
    -\int_{0}^{k_h} \rho^0\left(k\right) \mathrm{d}k \right].
    \label{excitedmomentumunpaired}
\end{equation}
We have  similar results for the one-particle-hole excitation of paired fermions, namely, 
\begin{equation}
    \begin{aligned}
        \Delta \Omega
        =\varepsilon^0\left(\lambda_p\right)-\varepsilon^0\left(\lambda_h\right), \quad
        \Delta K
        =2\pi\left[\int_{0}^{\lambda_p} \sigma^0\left(k\right) \mathrm{d}k
        -\int_{0}^{\lambda_h} \sigma^0\left(k\right) \mathrm{d}k \right].
    \end{aligned}
    \label{eqdispersionpaired}
\end{equation}
By giving all possible $k_p$ and $k_h$ ($\lambda_p$ and $\lambda_h$), one can obtain  one-particle-hole excitation of unpaired (paired) fermions. 
Accordingly, the spectrum can be plotted for certain choices of parameters $\{c,A,\mu_0H\}$, 
see Fig. \ref{1phboundary}. 

Apparently, there exists charge-charge separation at least for small $\Delta K$. 
The   particle-hole continuum  in the long-wavelength limit ($\Delta K\to 0$) manifests the free fermion-like dispersion. Two thresholds  of the excitation spectrum  (black solid lines in Fig. \ref{1phboundary}) reveal  a curvature effect. 
The first-order corrections to the  linear dispersions of paired  and unpaired fermions   can define the effective masses $m^*_b$ and $m^*_u$ as well as   the sound velocities, $v_b$ and $v_u$, see cyan and white dashed lines in Fig. \ref{1phboundary}, respectively. 
We  will  analytically calculate them in Section \ref{sectionapp}.
This is meaningful because it is much easier to measure specific signals referring to specific quantities, such as the sound velocities, which characterize a spectrum than to measure a whole spectrum in the experiment. 
By tuning the parameters $\{c,A,\mu_0H\}$, the ratio of sound velocities and Fermi surface of the two sectors are controllable, as shown in Fig. \ref{1ph}.

\begin{figure}[htbp]
    \centering
    \subfloat[ $1/\gamma$]{
        \includegraphics[width=0.48\textwidth]{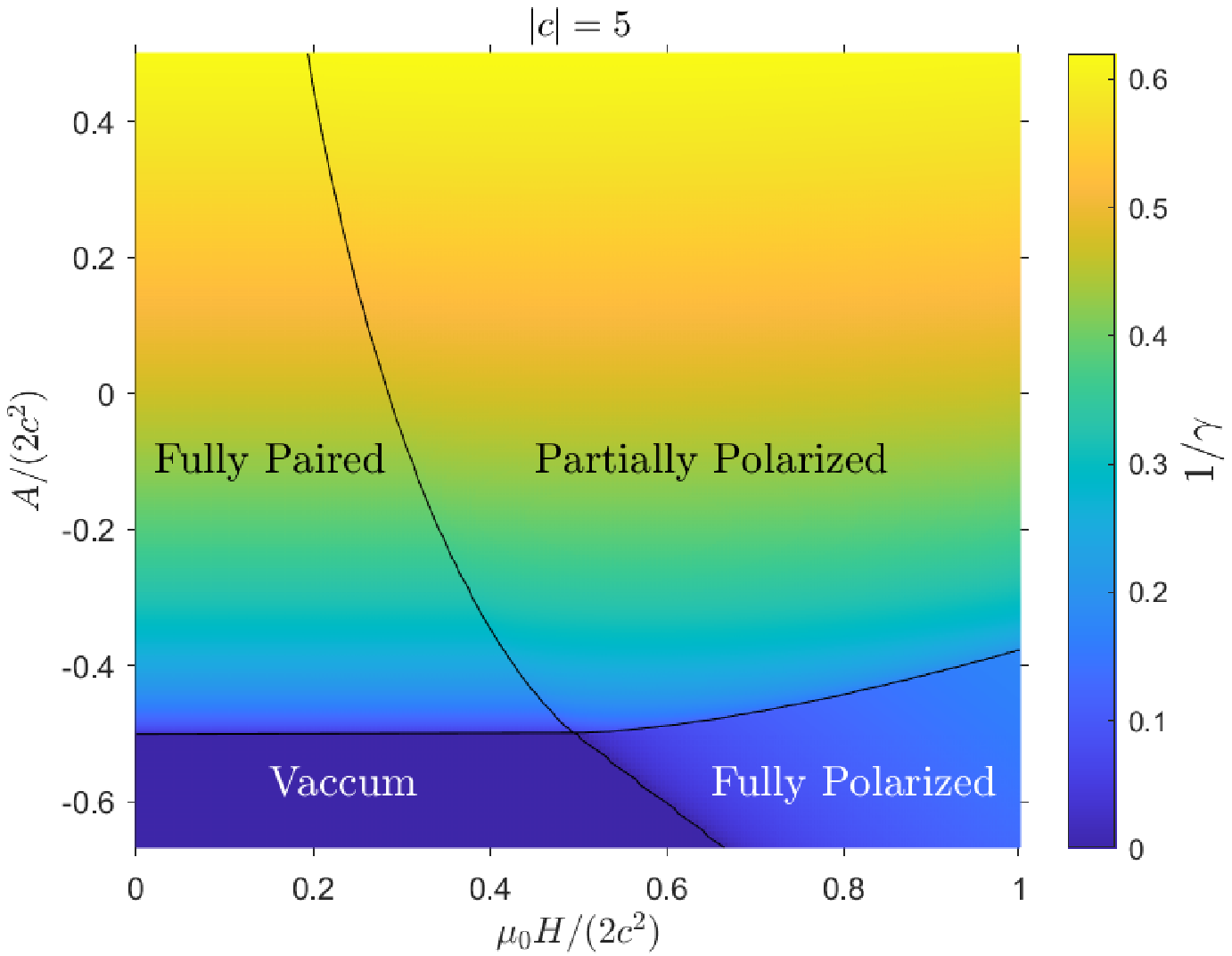} 
    }
    \subfloat[ $1/\gamma_b$ and $1/\gamma_u$]{
        \includegraphics[width=0.30\textwidth]{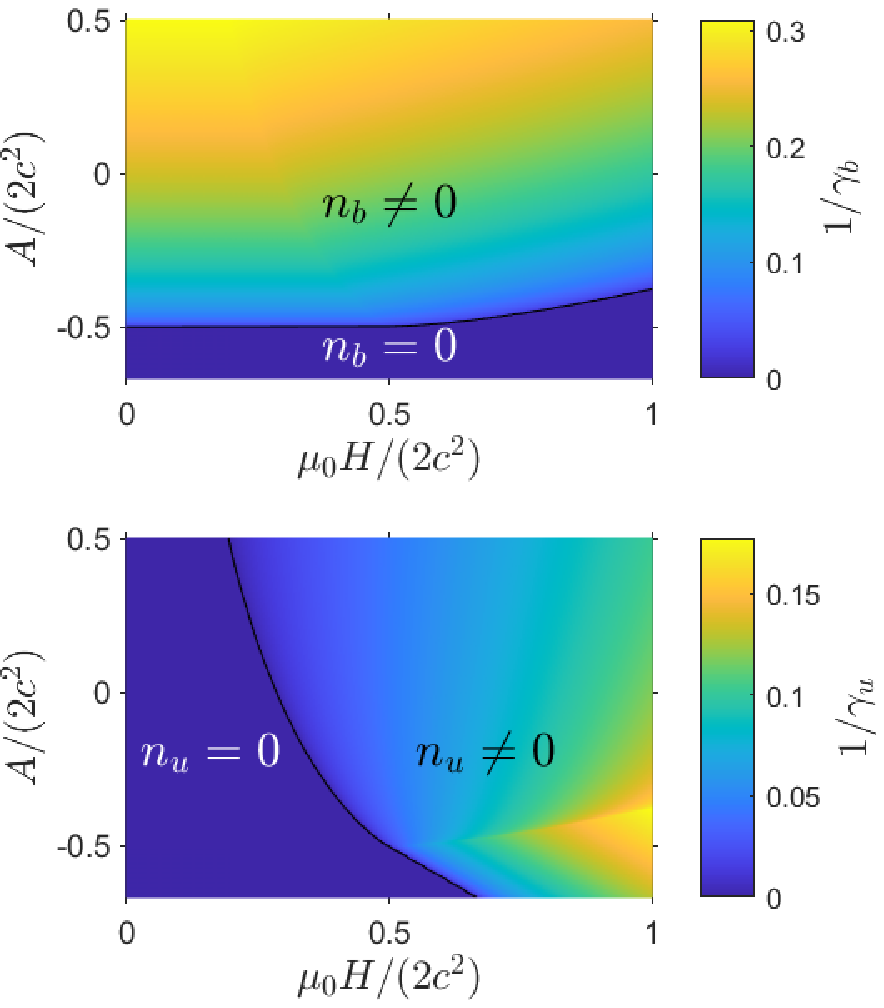} 
    }
    \caption[phase diagram represented by dimensionless particle densities $1/\gamma_b$, $1/\gamma_u$, and $1\gamma$]
    {\textbf{Phase diagrams represented by dimensionless particle densities $1/\gamma$, $1/\gamma_b$ and $1/\gamma_u$}.
    Here dimensionless quantities $A/(2c^2)$ and $\mu_0H/(2c^2)$ are used as variables, 
    where $2c^2$ is the binding energy.
    The black lines indicate the phase  boundaries, which are obtained from the two diagrams in (a). 
    For the partially polarized phase, the strong coupling $\gamma \gg 1$ near the quartet  point.     
 Clearly, according to (b), the boundaries of each phase (black lines) in (a) are given at the boundaries of zero- or nonzero-particle densities of paired and unpaired sections. The phase boundaries in (a) can also be determined by Eq. (\ref{eqdressed1}). 
    }
    \label{phasediagram_n}
\end{figure}

\section{Analytical Results}
\label{sectionapp}
\subsection{Strong Coupling Limit in Partially Polarized Phase}
\label{sectorstrongcoupling}
In general, analytical expressions of  physical quantities  are hardly obtained  from the BA equations unless in certain extreme limits. 
Usually, one consider the weak and strong coupling limits, $\left|c\right|\ll n$ and $\left|c\right|\gg n$ in 1D ultracold atomic systems. 
 The coupling strength $\left|c\right|/n$ can be experimentally tuned from weak to strong by controlling  the 3D scattering length near the Feshbach resonance, i.e. 
\begin{equation}
    \frac{c^2}{n^2}=\frac{mg_{1 \mathrm{D}}^2}{8\hbar^2n^2} \text{, and } 
    g_{1 \mathrm{D}}=
    \frac{2 \hbar^{2} a_{3 \mathrm{D}}}{m a_{\perp}^{2}} 
    \frac{1}{1-A a_{3 \mathrm{D}} / a_{\perp}},
\end{equation}
where $g_{1 \mathrm{D}}$ is the effective inter-component interaction, $a_{3 \mathrm{D}}$ is the 3D scattering length, $a_{\perp}$ is the transverse oscillator length.  
For our convenience, we denote the dimensionless coupling strengths $\gamma=\left|c\right|/n$, $\gamma_b=\left|c\right|/n_b$, and $\gamma_u=\left|c\right|/n_u$. 
What below  are the derivations of analytical results of physical quantities for the strong coupling regime, $\gamma\gg 1$.

\begin{figure}[htbp]
    \centering
    \subfloat[ polarization]{
        \includegraphics[width=0.42\textwidth]{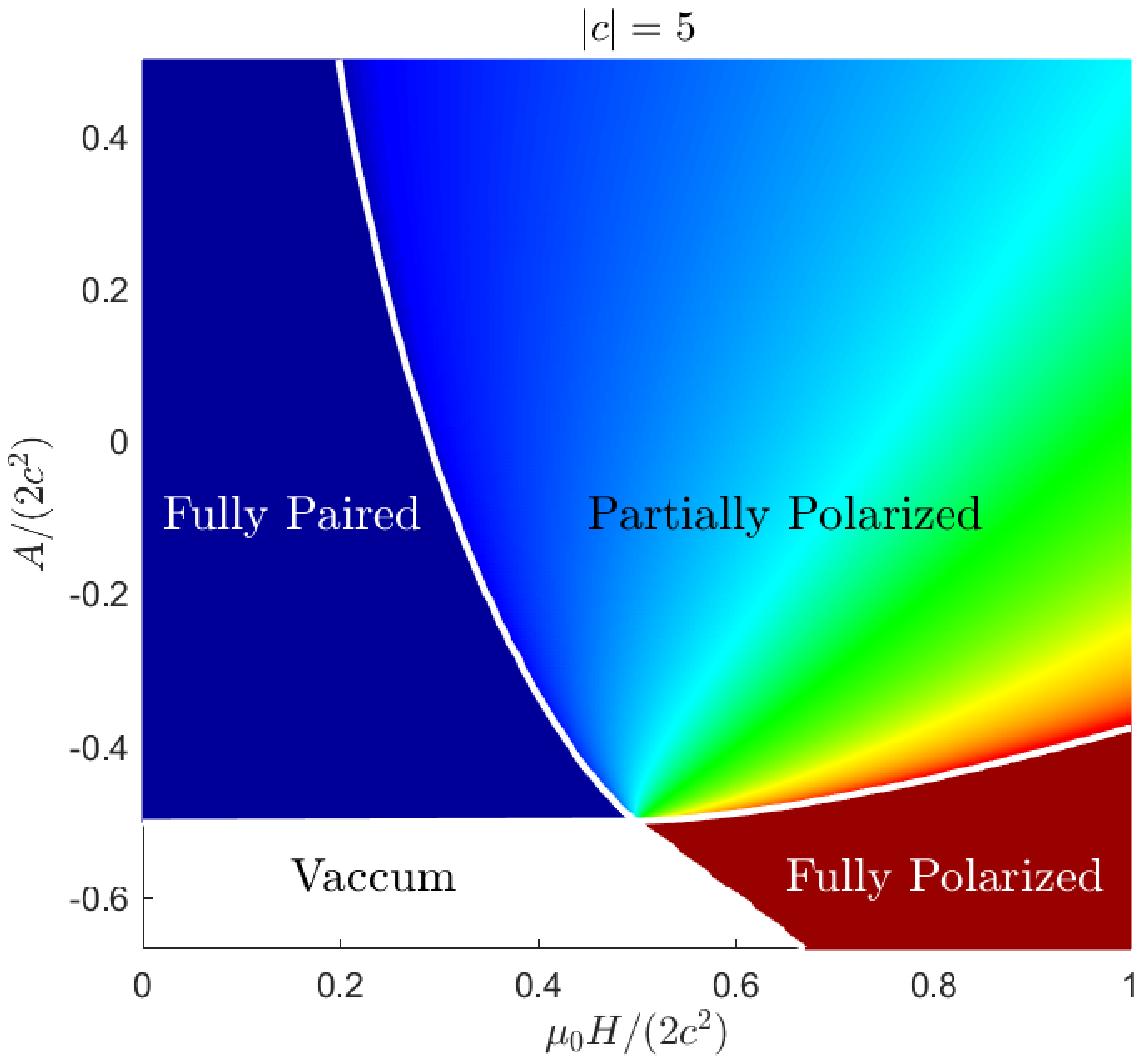} 
    }
    \subfloat[ polarization in 3D-version]{
        \includegraphics[width=0.48\textwidth]{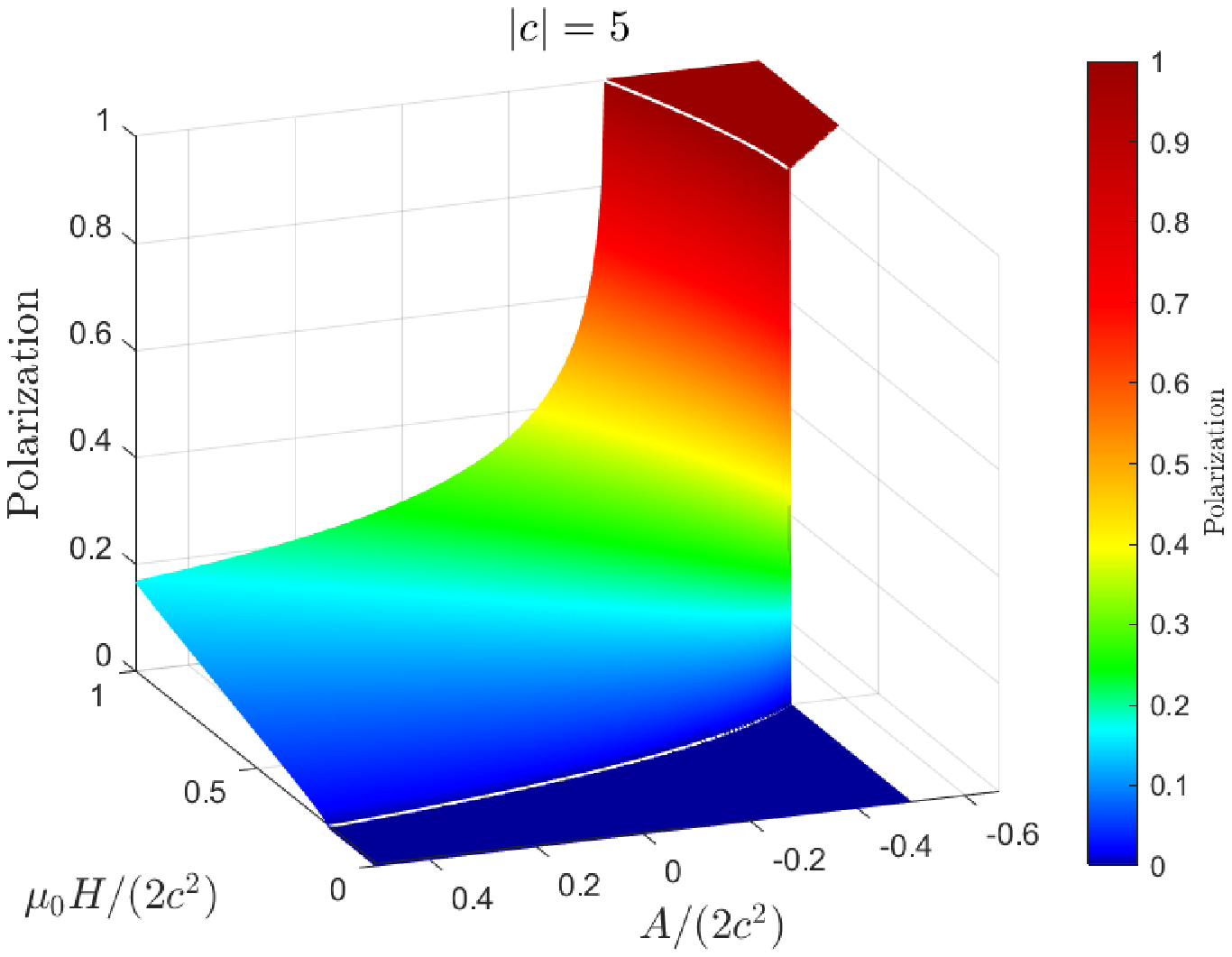} 
        \label{phasediagram_P_3D}
    }
    \caption[phase diagram of Polarization.]
    {\textbf{Phase diagram is represented by polarization.}
   In (a),  the white lines give the phase boundaries.
    In (b), The polarization varies drastically in the vicinity of the quartet point. 
    }
    \label{phasediagram_P}
\end{figure}

In the grand canonical ensemble,  quantum  phase transitions take place at zero-temperature, which have already been studied 
in  both the canonical ensemble and the grand canonical ensemble \cite{orso2007attractive,hu2007phase,guan2011quantum, yin2011quantum}. 
Here we discuss analytical results in the grand canonical ensemble, where the parameters $\{c,A,\mu_0H\}$ are our driving parameters.
For example, the phase digram and interaction strengths can be given in $\mu-H$ plane, see  Fig. \ref{phasediagram_n} and Fig. \ref{phasediagram_P}.

From  Fig. \ref{phasediagram_n}, we observe that the strong coupling $\gamma\gg 1$ (regions plotted in dark-blue) can be reached in the vicinity of the edge of the vacuum phase. 
In the partially polarized (FFLO-like) phase, the strong coupling regime  locates  near the quartet point  $A/c^2=-1$ and $\mu_0H/c^2=1$.
If we define the effective chemical potentials of both sector $A_b=A+c^2$ and $A_u=A+\mu_0H$, the strong coupling regime  meets the condition $A_b\ll c^2$ and $A_u\ll c^2$. 

In Fig. \ref{phasediagram_P} (a), we plot the polarization in the $\mu-H$ plane for fixed value of interaction strength.
From  \ref{phasediagram_P} (b), we observe the polarization $P=n_u/(2n_b+n_u)$ varies drastically in the vicinity of the quartet point.
Therefore, it might come into the expression of the band curvature corrections of sound velocities and effective masses for the strong coupling.

\subsection{Sound Velocity and Effective Mass in  One-Particle-hole Excitation Spectrum }
\label{sectionQB}
In this section, we will derive the exact dispersions of particle-hole excitations in both paired and unpaired fermions. 
We first derive the dispersion of bound pairs in the long-wavelength limit, i.e.,  
we derive the linear dispersion with a curvature correction. 
Consequently, we can obtain the sound velocities and effective masses in terms of dressed energies and the density distribution functions.
For a one-particle-hole excitation near the Fermi point  $B$ of the bound pairs, 
the excited energy and momentum are given by
\begin{equation}
    \begin{aligned}
        \Delta E_{b}
        =&\varepsilon\left(k_{p,i}\right)
        -\varepsilon\left(k_{h,i}\right)
        =\varepsilon\left(B+\Delta k\right)
        -\varepsilon\left(B\right), \\
        \Delta K_{b}
        =&2\pi\left[\int_{0}^{k_{p,i}} \sigma\left(k\right) \mathrm{d}k
        -\int_{0}^{k_{h,i}} \sigma\left(k\right) \mathrm{d}k \right]
        =2\pi\left[\int_{0}^{B+\Delta k} \sigma\left(k\right) \mathrm{d}k
        -\int_{0}^{B} \sigma\left(k\right) \mathrm{d}k \right].
    \end{aligned}
\end{equation}
In the long-wavelength limit,  the excitation is near the Fermi point, i.e., $\Delta k\ll 1$. 
Therefore, the excited energy and momentum can be expanded at $k=B$. Namely,
\begin{eqnarray}
        \Delta E_{b}
       & \approx&\varepsilon\left(B\right)+\varepsilon'\left(B\right)\Delta k
        +\frac{\varepsilon''\left(B\right)}{2}\left(\Delta k\right)^2
        =\varepsilon'\left(B\right)\Delta k
        +\frac{\varepsilon''\left(B\right)}{2}\left(\Delta k\right)^2, \\
        \Delta K_{b}
        &\approx&2\pi\int_{0}^{B} \sigma\left(k\right) \mathrm{d}k+2\pi\sigma\left(B\right)
        \Delta k
        +2\pi\frac{\sigma'\left(B\right)}{2}\left(\Delta k\right)^2
        -2\pi\int_{0}^{B} \sigma\left(k\right) \mathrm{d}k\nonumber\\
        &    =&2\pi\sigma\left(B\right)\Delta k
        +2\pi\frac{\sigma'\left(B\right)}{2}\left(\Delta k\right)^2, 
\end{eqnarray}
where $f'(x)$ and $f''(x)$ denote the first- and second-order derivatives of $f(x)$ with respect to $x$, respectively.
From the relation
$$
    \Delta K_{b}
    =2\pi\sigma\left(B\right)\Delta k
    \left(1+\frac{\sigma'\left(B\right)}{2\sigma\left(B\right)}\Delta k\right),
$$
we directly obtain $\Delta k$,   
\begin{eqnarray}
    \Delta k
    \approx
   \frac{1}{2\pi\sigma(B)}\Delta K_{b}
    -\frac{\sigma'\left(B\right)}{8\pi^2\left(\sigma\left(B\right)\right)^3}
    \left(\Delta K_{b}\right)^2.
\end{eqnarray}
Subsequently, we have  the following dispersion relation for the bound pairs 
\begin{equation}
    \begin{aligned}
        \Delta E_{b}
        &\approx \varepsilon(B)\left(
        \frac{1}{2\pi\sigma(B)}\Delta K_{b}
        -\frac{\sigma'\left(B\right)}{8\pi^2\left(\sigma\left(B\right)\right)^3}
        \left(\Delta K_{b}\right)^2\right)
        +\frac{\varepsilon''(B)}{2}\left(\frac{1}{2\pi\sigma(B)}\Delta K_{b}\right)^2\\
        &\approx \frac{\varepsilon'(B)}{2\pi\sigma(B)}\Delta K_{b}
        +\left[\frac{\varepsilon''(B)}{8\pi^2\left(\sigma(B)\right)^2}
        -\frac{\sigma'(B)\varepsilon'(B)}{8\pi^2\left(\sigma(B)\right)^3}\right]
        (\Delta K_{b})^2\\
        &\approx 2\left[v_{b} \frac{\Delta K_{b}}{2}
        +\frac{1}{2 m^{*}_b}\left(\frac{\Delta K_{b}}{2}\right)^{2}\right]
    \end{aligned}
    \label{exenergyinexmonentum}
\end{equation}
expressed up to the second-order of $\Delta K_b$. 
From this relation, we read off the sound velocity and effective mass of the one-particle-hole excitation in the paired Fermi sea as
\begin{equation}
    \begin{aligned}
        v_{b}
        &=\frac{\varepsilon'(B)}{2\pi\sigma(B)},\\
        \frac{1}{4 m^{*}_b}
        &=\frac{\varepsilon''(B)}{8\pi^2\left(\sigma(B)\right)^2}
        -\frac{\sigma'(B)\varepsilon'(B)}{8\pi^2\left(\sigma(B)\right)^3},
    \end{aligned}
    \label{appendixvcmintermpaired}
\end{equation}
which hold true for arbitrary coupling strength and thus will be used in our numerical calculations.

Similarly, a one-particle-hole excitation for unpaired fermions  near the Fermi point can be calculated via the excited energy and momentum,  
\begin{equation}
    \begin{aligned}
        \Delta E_u
        =&\left[\kappa\left(k_{p,i}\right)-\kappa\left(k_{h,i}\right)\right]
        =\left[\kappa\left(Q+\Delta k\right)-\kappa\left(Q\right)\right] 
        \approx \kappa'\left(Q\right)\Delta k
        +\frac{\kappa''\left(Q\right)}{2}\left(\Delta k\right)^2, \\
        \Delta K_u
        =&2\pi\left[\int_{0}^{k_{p,i}} \rho\left(k\right) \mathrm{d}k
        -\int_{0}^{k_{h,i}} \rho\left(k\right) \mathrm{d}k \right]
        =2\pi\left[\int_{0}^{Q+\Delta k} \rho\left(k\right) \mathrm{d}k
        -\int_{0}^{Q} \rho\left(k\right) \mathrm{d}k \right]\\
        \approx& 2\pi\rho\left(Q\right)\Delta k
        +2\pi\frac{\rho'\left(Q\right)}{2}\left(\Delta k\right)^2.
    \end{aligned}
\end{equation}
The excited energy can be expressed by the excited momentum as well as 
Eq.(\ref{exenergyinexmonentum}), 
\begin{equation}
    \begin{aligned}
        \Delta E_{u}
        &=\kappa(Q)\left(
        \frac{1}{2\pi\rho(Q)}\Delta K_{u}
        -\frac{\rho'(Q)}{8\pi^2\left(\rho(Q)\right)^3}
        \left(\Delta K_{u}\right)^2\right)
        +\frac{\kappa''(Q)}{2}\left(\frac{1}{2\pi\rho(Q)}\Delta K_{u}\right)^2\\
        &=\frac{\kappa'(Q)}{2\pi\rho(Q)}\Delta K_{u}
        +\left[\frac{\kappa''(Q)}{8\pi^2\left(\rho(Q)\right)^2}
        -\frac{\rho'(Q)\kappa'(Q)}{8\pi^2\left(\rho(Q)\right)^3}\right]
        (\Delta K_{u})^2
        =v_{u} \Delta K_{u}
        +\frac{1}{2 m^{*}_b}\left(\Delta K_{u}\right)^{2}.
    \end{aligned}
\end{equation}
Therefore, by comparison, the sound velocity and effective mass of  the one-particle-hole excitation in the unpaired Fermi sea are given by
\begin{equation}
    \begin{aligned}
        v_{u}
        &=\frac{\kappa'(Q)}{2\pi\rho(Q)},\\
        \frac{1}{2 m^{*}_u}
        &=\frac{\kappa''(Q)}{8\pi^2\left(\rho(Q)\right)^2}
        -\frac{\rho'(Q)\kappa'(Q)}{8\pi^2\left(\rho(Q)\right)^3}.
    \end{aligned}
    \label{appendixvcmintermunpaired}
\end{equation}
The obtained relations for velocities and effective masses (\ref{appendixvcmintermpaired}) and (\ref{appendixvcmintermunpaired}) are very convenient to carry out numerical calculations, see Fig. \ref{1phboundary}.

For a strong coupling, i.e. $|c|\gg 1$, we may directly use the dressed energies Eq.(\ref{eqdressed1}) and the distribution functions Eq.(\ref{eqdistribution}) to determine the dispersion relations characterized by the sound velocities and effective masses. 
To this end, we first calculate Fermi momenta $B$ and $Q$ in terms of particle densities $n_u$, $n_b$ with  our chose parameters $\{c,A,\mu_0H\}$ (i.e., $[A_b, A_u]$). 
For the strong coupling, we may expand the kernels $a_1(k)$ and $a_2(k)$  up to the first order of $1 /c$, the distribution functions Eq.(\ref{eqdistribution}) thus  become
\begin{equation}
    \begin{aligned}
        &\sigma\left(k\right)\approx\frac{1}{\pi}
        - a_2\left(k\right)\int_{-B}^{B} \sigma\left(k'\right)\mathrm{d}k'
        - a_1\left(k\right)\int_{-Q}^{Q} \rho\left(k'\right)\mathrm{d}k'
        = \frac{1}{\pi}-a_2\left(k\right)n_b - a_1\left(k\right)n_u, \\
        &\rho\left(k\right)\approx\frac{1}{2 \pi}
        - a_1\left(k\right)\int_{-B}^{B} \sigma\left(k'\right)\mathrm{d}k'
        = \frac{1}{2 \pi}-a_1\left(k\right)n_b.
    \end{aligned}
    \label{approxdistribution}
\end{equation}
The paiticle densities $n_b$ and $n_u$ are related to  Fermi momenta $B$ and $Q$ via 
\begin{equation}
    \begin{aligned}
        n_b&=
        \int_{-B}^{B} \sigma\left(k\right)\mathrm{d}k
        \approx \int_{-B}^{B} 
        \left(\frac{1}{\pi}-a_2\left(k\right)n_b - a_1\left(k\right)n_u\right)
        \mathrm{d}k\\
        &=\frac{2B}{\pi}
        -\frac{2}{\pi}\arctan{\frac{B}{2\left|c\right|}}n_b
        -\frac{2}{\pi}\arctan{\frac{B}{\left|c\right|}}n_u 
        \approx\frac{2B}{\pi}\left(1-\frac{B}{\pi \left|c\right|}
        -\frac{Q}{\pi \left|c\right|}\right), \\
        n_u&=
        \int_{-Q}^{Q} \rho\left(k\right)\mathrm{d}k
        \approx \int_{-Q}^{Q} \left(\frac{1}{2 \pi}-a_1\left(k\right)n_b\right)\mathrm{d}k
        =\frac{Q}{\pi} -\frac{2}{\pi}\arctan{\frac{Q}{\left|c\right|}}n_b
        \approx\frac{Q}{\pi}\left(1-\frac{4B}{\pi \left|c\right|}\right),
    \end{aligned}
    \label{eqnbnuinBQ}
\end{equation}
which  suggest the following   relations  between particle densities ($n_u$ and $n_b$) and Fermi momenta ($B$ and $Q$)
\begin{equation}
    \begin{aligned}
        &B=\frac{1}{2}\pi n_b \left(1-\frac{B}{\pi \left|c\right|}
        -\frac{Q}{\pi \left|c\right|}\right)^{-1}
        \approx \frac{1}{2}\pi n_b \left(1+\frac{B+Q}{\pi \left|c\right|}\right), \\
        &Q=\pi n_u \left(1-\frac{4B}{\pi \left|c\right|}\right)^{-1}
        \approx \pi n_u \left(1+\frac{4B}{\pi \left|c\right|}\right).
    \end{aligned}
    \label{BQinnbnu}
\end{equation}
For a further derivation of the velocities, we need  the effective chemical  pressures of pairs and unpaired fermions
\begin{equation}
    \begin{aligned}
        p_b=-\frac{1}{\pi}\int_{-B}^{B}\varepsilon(k)\mathrm{d}k,\quad
        p_u=-\frac{1}{2\pi}\int_{-Q}^{Q}\kappa(k)\mathrm{d}k. 
    \end{aligned}
    \label{sectionpressure}
\end{equation}
From the dressed energy  Eq.(\ref{eqdressed1}), we may obtain the following relations between effective chemical potentials and Fermi momenta \cite{guan2007phase}
\begin{equation}
    \begin{aligned}
        B^2\approx A+c^2-\frac{p_b+4p_u}{4\left|c\right|}=A_b-\frac{p_b+4p_u}{4\left|c\right|},\quad
        Q^2\approx A+\mu_0H-\frac{p_b}{\left|c\right|}=A_u-\frac{p_b}{\left|c\right|}.
    \end{aligned}
    \label{BQineffectivechemicalpotential}
\end{equation}
Consequently, according to Eqs.(\ref{BQinnbnu}) and (\ref{BQineffectivechemicalpotential}),  we built up the relations between the particle densities  and our chosen parameters $\{c,A,\mu_0H\}$ through the Fermi point $B$ and $Q$. 

Now we can rewrite the expressions of the sound velocities and effective masses  following the expressions Eq.(\ref{appendixvcmintermpaired}) and (\ref{appendixvcmintermunpaired}) for the strong coupling regime. 
To determine $v_b$ and $m_b^*$ given by Eq.(\ref{appendixvcmintermpaired}),  we first calculate $\varepsilon'(B)$ and $\sigma(B)$ and their derivatives. 
According to Eq.(\ref{eqdressed1}),   $\varepsilon'(B)$ is given by 
\begin{equation}
    \varepsilon'\left(B\right)=4B
    - \int_{-B}^{B} \left.\frac{\mathrm{d}a_2\left(k-k'\right)}{\mathrm{d}k}\right|_{k=B}\varepsilon\left(k'\right)\mathrm{d}k'
    - \int_{-Q}^{Q} \left.\frac{\mathrm{d}a_1\left(k-k'\right)}{\mathrm{d}k}\right|_{k=B}\kappa\left(k'\right)\mathrm{d}k' 
    \approx 4B, 
\end{equation}
where we ignored  the order $1/c^2$, for example,   the leading order of  $\mathrm{d}a_n(k)/\mathrm{d}k$ is $\mathcal{O}(1/c^3)$ in  the strong coupling limit. 
Similarly, we can have  $\varepsilon''\left(B\right)=4+ \mathcal{O}(1/c^4)$. 
Whereas the distribution function $\sigma\left(B\right)$ can be obtained by substituting Eqs.(\ref{approxdistribution}) into Eq.(\ref{eqdistribution}), namely, 
\begin{equation}
    \begin{aligned}
        &\sigma\left(B\right)\\=&\frac{1}{\pi}
        - \int_{-B}^{B} a_2\left(B-k'\right)
        \left(\frac{1}{\pi}-a_2\left(k'\right)n_b - a_1\left(k'\right)n_u\right)
        \mathrm{d}k'
        - \int_{-Q}^{Q} a_1\left(B-k'\right)
        \left(\frac{1}{2 \pi}-a_1\left(k'\right)n_b\right)
        \mathrm{d}k'\\
        \approx&\frac{1}{\pi}
        - \frac{1}{\pi}\int_{-B}^{B} \frac{1}{2\pi\left|c\right|}\mathrm{d}k'
        - \frac{1}{2 \pi}\int_{-Q}^{Q} \frac{1}{\pi\left|c\right|}\mathrm{d}k'
        \approx\frac{1}{\pi}
        \left(1- \frac{B+Q}{\pi\left|c\right|}\right).
    \end{aligned}
\end{equation}
While its derivative $\sigma'\left(B\right)\sim\mathcal{O}\left(1/c^3\right)$ is negligible. 
By substituting $\varepsilon'(B)$ and $\sigma(B)$ and their derivatives into Eq.(\ref{appendixvcmintermpaired}), we may express $v_b$ and $m_b^*$ with respect to Fermi momenta $B$ and $Q$
\begin{equation}
    \begin{aligned}
        &v_{b}
        =\frac{\varepsilon'(B)}{2\pi\sigma(B)}
        =2B\left(1- \frac{B+Q}{\pi\left|c\right|}\right)^{-1}
        \approx 2B\left(1+ \frac{B+Q}{\pi\left|c\right|}\right),\\
        &\frac{1}{4 m^{*}_b}
        =\frac{\varepsilon''(B)}{8\pi^2\left(\sigma(B)\right)^2}
        -\frac{\sigma'(B)\varepsilon'(B)}{8\pi^2\left(\sigma(B)\right)^3}
        \approx \frac{1}{2}\left(1- \frac{B+Q}{\pi\left|c\right|}\right)^{-2}.
    \end{aligned}
    \label{appendixvcmBQ}
\end{equation}
By substituting Eqs.(\ref{BQinnbnu}) into the last equations, we further express $v_b$ and $m_b^*$ in $\gamma_b$ and $\gamma_u$ 
\begin{equation}
    \begin{aligned}
        v_{b}
        &\approx \pi n_b\left(1+ \frac{B+Q}{\pi\left|c\right|}\right)^2 
        \approx \pi n_b\left(1+ \frac{2B+2Q}{\pi\left|c\right|}\right)
        \approx \pi n_{b}\left(1+\frac{1}{\gamma_b}+\frac{2}{\gamma_u}\right),\\
        m^{*}_b
        &\approx \frac{1}{2}\left(1- \frac{B+Q}{\pi\left|c\right|}\right)^{-2}
        \approx \frac{1}{2}\left(1+ \frac{2B+2Q}{\pi\left|c\right|}\right)
        = m_b\left(1+ \frac{2B+2Q}{\pi\left|c\right|}\right)
        \approx m_b\left(1-\frac{1}{\gamma_b}-\frac{2}{\gamma_u}\right).
        \label{vbmb}
    \end{aligned}
\end{equation} 

Similarly, we may obtain  $v_u$ and $m_u^*$ from  Eq.(\ref{appendixvcmintermunpaired}).
The terms $\kappa'(Q)$ and $\rho(Q)$  are given by 
\begin{equation}
    \begin{aligned}
        &\kappa'(Q)=2Q,&
        \kappa''(Q)&=2,\\
        &\rho(Q)=\frac{1}{2\pi}\left(1- \frac{4B}{\pi\left|c\right|}\right),\quad&
        \rho'(Q)&\sim\mathcal{O}\left(\frac{1}{c^3}\right)
    \end{aligned}
\end{equation}
up to the order   of $1/c^2$. 
Substituting them into Eqs.(\ref{appendixvcmintermunpaired}), we obtain  $v_u$ and $m_u^*$ in the following forms 
\begin{equation}
    \begin{aligned}
        v_{u}
        &\approx 2\pi n_u\left(1+ \frac{4B}{\pi\left|c\right|}\right)^2 
        \approx 2\pi n_u\left(1+ \frac{8B}{\pi\left|c\right|}\right)
        \approx 2\pi n_{u}\left(1+\frac{4}{\gamma_b}\right),\\
        m^{*}_u
        &\approx \left(1- \frac{4B}{\pi\left|c\right|}\right)^{-2}
        \approx \left(1+ \frac{8B}{\pi\left|c\right|}\right)
        = m\left(1+ \frac{8B}{\pi\left|c\right|}\right)
        \approx m\left(1-\frac{4}{\gamma_b}\right).
    \end{aligned}
    \label{vumu}
\end{equation}

\begin{figure}[htbp]
    \centering
    \includegraphics[width=0.95\textwidth]{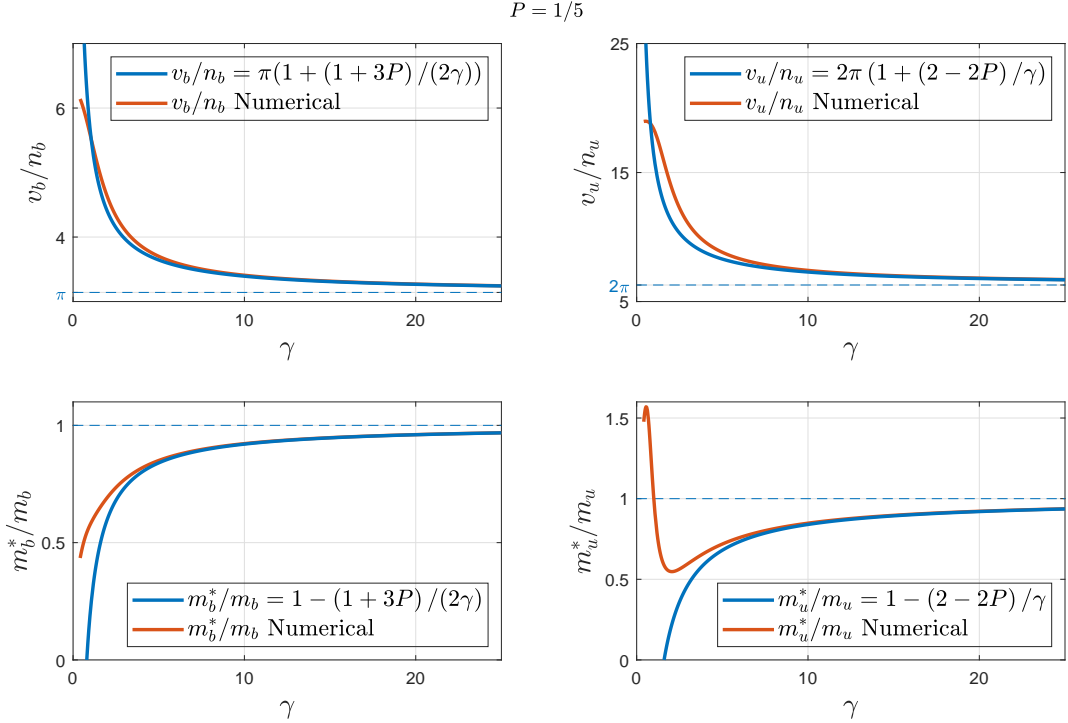} 
    \caption[variation of sound velocities and effective masses 
    in dimensionless interaction strengths $\gamma$.]
    {\textbf{Variation of sound velocities, $v_{b}$ and $v_{u}$, and effective masses, $m^*_b$ and $m^*_u$ v.s. the  dimensionless interaction strengths $\gamma$.}
    The polarization is fixed to be $P=1/5$ with  $\gamma=10$.
    The analytical (blue) Eq.(\ref{vcmingamma}) are agreeable with the numerical (red) results which are plotted with respect to  Eqs.(\ref{appendixvcmintermpaired}) and (\ref{appendixvcmintermunpaired}). 
    }
    \label{vm-gamma}
\end{figure}

Furthermore, owing to Eqs.(\ref{BQinnbnu}) and (\ref{BQineffectivechemicalpotential}), we  replace $\gamma_b$ and $\gamma_u$ in Eqs.(\ref{vbmb}) and (\ref{vumu}) by $\gamma$ and the polarization $P$ or the ratio between two effective chemical potentials $A_b$ and $A_u$. 
Since the polarization is defined by $P=n_u/n$, we have $\gamma/\gamma_b=n_b/n=(1-P)/2$ and $\gamma/\gamma_u=n_u/n=P$, which allows us to express the sound velocities and effective masses  as a function of $P$. 
Alternatively, we would like to express the sound velocities and effective masses in terms of $A_b$ and $A_u$. Since they are directly related to our controllable parameters $\{c,A,\mu_0H\}$. 
According to Eqs.(\ref{BQinnbnu}) and (\ref{BQineffectivechemicalpotential}), we have
\begin{equation}
    \begin{aligned}
        \left(\frac{Q}{B}\right)^2
        \approx\left(\frac{2n_u}{n_b}\right)^2
        =4\left(\frac{2P}{1-P}\right)^2
        \approx\frac{A_u}{A_b}
        =\alpha^2. 
    \end{aligned}
    \label{eqnu/nb}
\end{equation}
We  denote the ratio of effective chemical potential as $A_u/A_b=\alpha^2$ for our  convenience in analysis. 
Using the polarization  $P$ and the ratio $\alpha$,  we can rewrite Eqs.(\ref{vbmb}) and (\ref{vumu}) as
\begin{equation}
    \begin{aligned}
        &v_{b}
        \approx\pi n_{b}\left(1+\left(1+3P\right)\frac{1}{2\gamma}\right)
        \approx\pi n_{b}\left(1+\frac{2+2\alpha}{4+\alpha}\frac{1}{\gamma}\right), \\
        &v_{u}
        \approx 2\pi n_{u}\left(1+\left(2-2P\right)\frac{1}{\gamma}\right)
        \approx 2\pi n_{u}\left(1+\frac{8}{4+\alpha}\frac{1}{\gamma}\right),\\
        &m^{*}_b
        \approx m_b\left(1-\left(1+3P\right)\frac{1}{2\gamma}\right)
        \approx m_b\left(1-\frac{2+2\alpha}{4+\alpha}\frac{1}{\gamma}\right), \\
        &m^{*}_u
        \approx m_u\left(1-\left(2-2P\right)\frac{1}{\gamma}\right)
        \approx m_u\left(1-\frac{8}{4+\alpha}\frac{1}{\gamma}\right), 
    \end{aligned}
    \label{vcmingamma}
\end{equation}
which are confirmed by  numerical results in Fig. \ref{vm-gamma},  where we set the  polarization $P=1/5$ (i.e., $\alpha=1$).
According to Eq.(\ref{eqnu/nb}) and the last equations Eqs (\ref{vcmingamma}), the ratio between two sound velocities is simply given by $v_u/v_b=2n_u/n_b=\alpha$ up to the order of $1/c^2$, which provides physical insight into  the sound velocities in  Fig. \ref{1ph}, where our chosen parameters $\{c,A,\mu_0H\}$ indicate a strong coupling  with $\alpha=2$ and $\alpha=1$, respectively.
Furthermore, since $k_{F,b}\approx\pi n_b$ and $k_{F,u}\approx\pi n_u$ (illustrated at the end of Section \ref{sectiongroundstate}), we have $k_{F,u}/k_{F,b}=\alpha/2$, which gives a special  ratio between two Fermi points in Fig. \ref{1ph}.

So far, we derived the sound velocities and effective masses of paired and unpaired fermions starting from the zero-temperature TBA equations. 
For the repulsive case, it is proved\cite{he2020emergence} that these quantities can be derived from the BA equations; by a similar method, one can prove that this works for the attractive case as well (See Appendix.\ref{appendixvcm} for detailed procedure). 
This is reasonable since the TBA equations are derived from the BA equations.

The sound velocities and effective masses can effectively characterize the spectra of one-particle-hole excitation. 
Therefore, these quantities can be regarded as evidence and provide convenience to varify the separation of collective motions within paired and unpaired charges at zero temperature. 
This kind of charge-charge separation can even be characterized at non-zero but low temperature. 
According to the relations given by Eqs.(\ref{appendixvcmintermpaired}) and (\ref{appendixvcmintermunpaired}), the zero-temperature subtraction for low-temperature free energy density can be expressed by (see Appendix.\ref{Free Energy at Low-Temperature} for proof)
\begin{equation}
    \begin{aligned}
        \Omega/L-\Omega_0/L=p-p_0
        =\frac{\pi T^2}{6}\left(\frac{1}{v_b}+\frac{1}{v_u}\right),
    \end{aligned}
    \label{eqpressureinvelocity}
\end{equation}
which is simply the addition of contribution of paired and unpaired sections. 
This additivity manifests the independent character and separation of paired and unpaired bosonic modes. Moreover, the $T^2$ dependence given by the low-temperature pressure correction shows a typical linear dispersions feature. The specific heat can also be obtained by its definition, i.e., 
\begin{equation}
    \begin{aligned}
        c_{V}=T \frac{\partial^2 p}{\partial T^2}
        =\frac{\pi T}{3}\left(\frac{1}{v_b}+\frac{1}{v_u}\right),
    \end{aligned}
    \label{eqspecificheatinvelocity}
\end{equation}
which preserves the additivity and describes the novel charge-charge separation as well at low temperature.

\section{Excitations other than single  particle-hole }
\label{sectionother}

In this section, we will consider excitations other than the one particle-hole excitation, such as excitations of multiple particle-holes, breaking and forming pairs, as well as  length-$n$ string excitations. 

\subsection{Multiple particle-hole Excitations}

\begin{figure}[htbp]
    \centering
    \subfloat[$N_b=2$, $N_u=0$]{
    \includegraphics[width=0.3\textwidth]{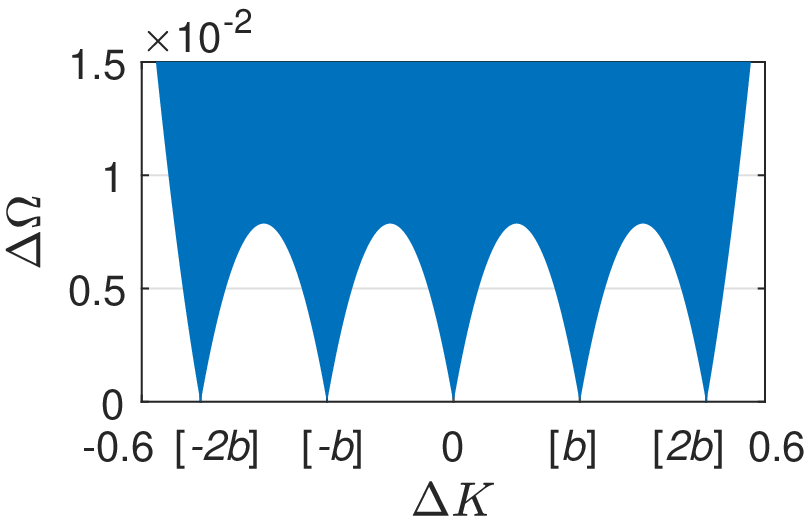}
    }
    \subfloat[$N_b=0$, $N_u=2$]{
    \includegraphics[width=0.3\textwidth]{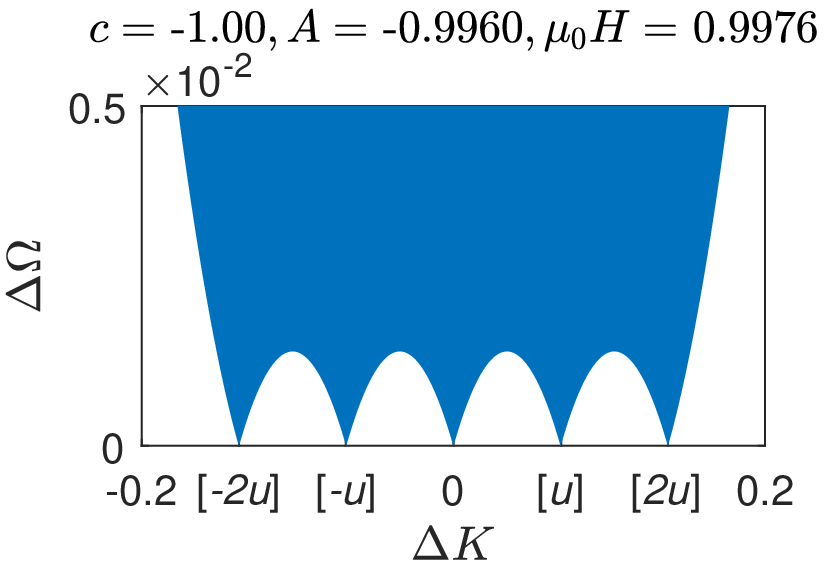}
    }
    \subfloat[$N_b=1$, $N_u=1$]{
    \includegraphics[width=0.35\textwidth]{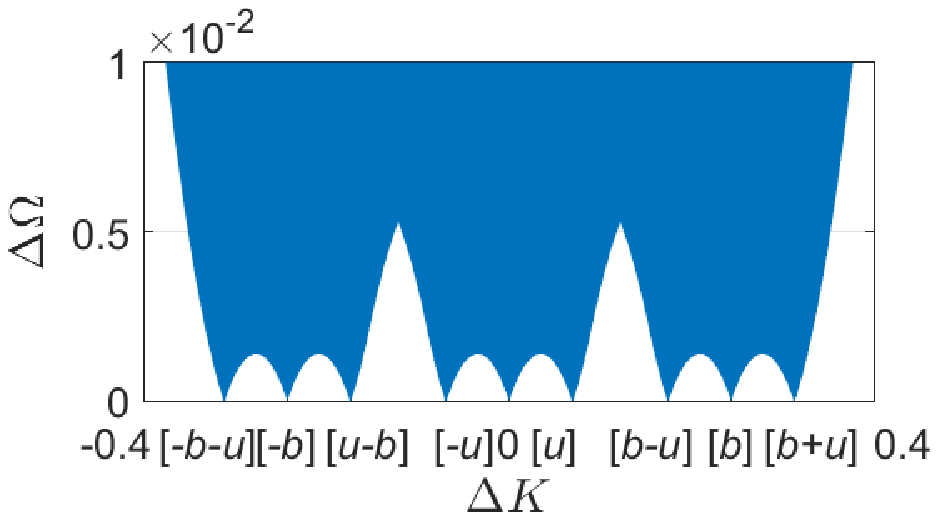}
    }
    \caption[Multiple particle-hole excitations spectra]
    {\textbf{Two-particle-hole excitations spectra}.
    The numbers of excited particle-holes in paired and unpaired sectors.   The figures (a), (b) and (c) show particle-hole excitations for different $N_b$ and $N_u$. 
Here we denote  $[\alpha b+\beta u]\equiv: 2\left(\alpha k_{F,b}+\beta k_{F,u}\right)$. 
    In this figure, we set $c=-1.00$, $A=-0.9960$, $\mu_0H=0.9976$ for our numerical calculation with the dispersions represented by Eq.(\ref{dispersionmulti})
    }
    \label{phb2}
\end{figure}

In the case of exciting $N_b$ paired fermions and $N_u$ unpaired fermions, like what  we did in Eq.(\ref{distributionintroducedelta}),  introducing $2N_b+2N_u$ $\delta$-functions, $\sum_{i=1}^{N_b}[-\delta\left(k-k_{h,i}\right)+\delta\left(k-k_{p,i}\right)]/L+\sum_{j=1}^{N_u}[-\delta\left(k-\right.$ $\left.k_{h,j}\right)+\delta\left(k-k_{p,j}\right)]/L$ into the distribution functions Eq.(\ref{eqdistribution}), one can  prove that the excitation energy  momentum are   given by  their summation of  one-particle-hole excitation ones, respectively 
\begin{equation}
    \begin{aligned}
        \Delta \Omega
        =&\sum_{i=1}^{N_b}
        \left[\varepsilon^0\left(\lambda_{p,i}\right)
        -\varepsilon^0\left(\lambda_{h,i}\right)\right]
        +\sum_{i=1}^{N_u}
        \left[\kappa^0\left(k_{p,i}\right)-\kappa^0\left(k_{h,i}\right)\right] \\
        \Delta K
        =&2\pi\sum_{i=1}^{N_b}
        \left[\int_{0}^{\lambda_{p,i}} \sigma^0\left(k\right) \mathrm{d}k
        -\int_{0}^{\lambda_{h,i}} \sigma^0\left(k\right) \mathrm{d}k \right]
        +2\pi\sum_{i=1}^{N_u}\left[\int_{0}^{k_{p,i}} \rho^0\left(k\right) \mathrm{d}k
        -\int_{0}^{k_{h,i}} \rho^0\left(k\right) \mathrm{d}k \right].
        \label{dispersionmulti}
    \end{aligned}
\end{equation}
The calculation is essentially similar as that for one-particle-hole dispersion. Here  we do not wish to present the calculation in detail. 
 Fig. \ref{phb2} shows three examples for  various choices of the numbers  $N_b$ and $N_u$.

\subsection{Excitations by pairing and depairing without exciting any n-string}
\label{sectionbreakingforming}

\begin{figure}[htbp] 
    \centering
    \def\svgwidth{\columnwidth}
    \includegraphics[width=0.65\textwidth]{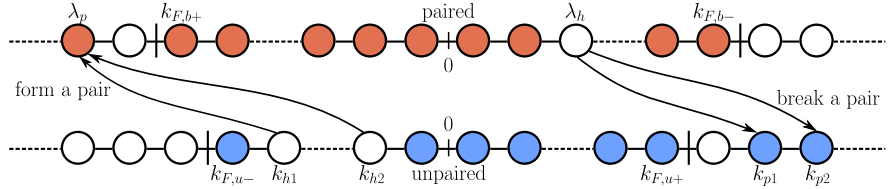}
    \caption[sketch of forming or breaking a pair.  ]
    {\textbf{Schematic illustration of forming or breaking a pair}. 
    }
    \label{figbreakpair}
\end{figure}

\begin{figure}[htbp]
    \centering
    \subfloat[breaking one pair]{
    \includegraphics[width=0.48\textwidth]{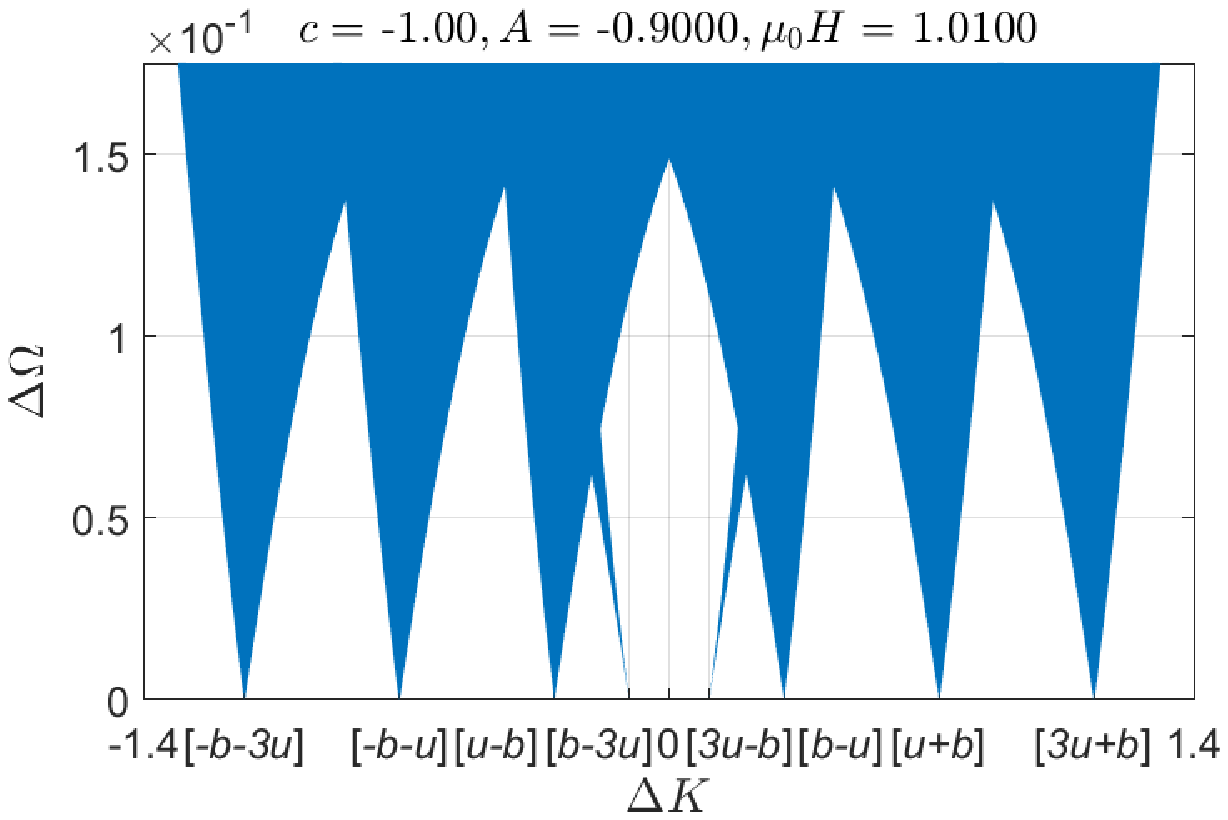}
    }
    \subfloat[forming one pair]{
    \includegraphics[width=0.48\textwidth]{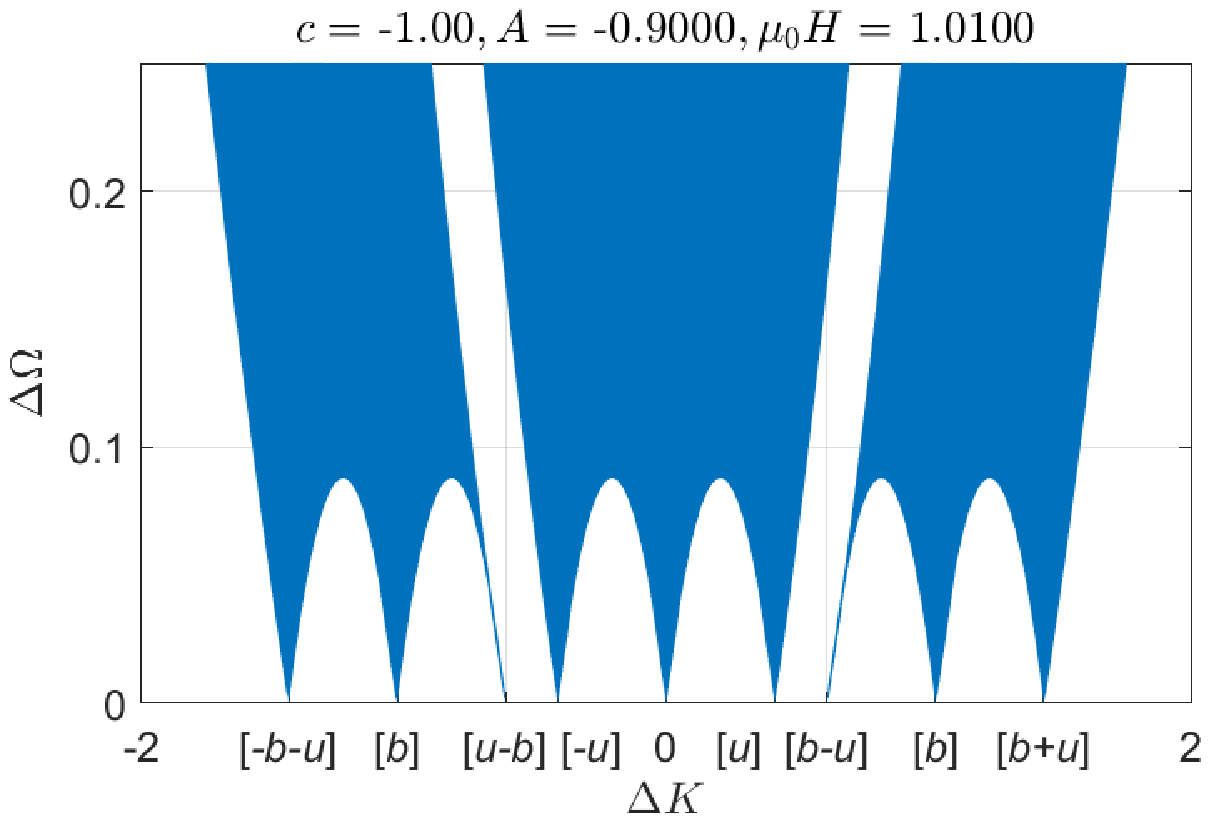}
    }
    \caption[excited spectrum for breaking one pair]
    {\textbf{Excited spectrum of breaking or forming one pair}.
    The excitation spectra show  two-fold degeneracy in breaking a pair and forming a pair. 
  Here we denote  $[\alpha b+\beta u] \equiv: 2\left(\alpha k_{F,b}+\beta k_{F,u}\right)$. 
   We  set $c=-1.00$, $A=-0.9000$, $\mu_0H=1.0100$ for our numerical calculation with the dispersions Eqs.(\ref{excitedenergybreakingforming}) and (\ref{excitedmomentumbreakingforming}).
    }
    \label{Nub}
\end{figure}

For the case of breaking $N_{ub}$ pairs, we create   $N_{ub}$ holes in the paired Fermi sea while add  $2N_{ub}$ fermions in the unpaired sea, see Fig.~\ref{figbreakpair}. 
For  the case of forming $N_{bu}$ pairs, we create $2N_{bu}$ holes in the unpaired Fermi sea and add $N_{bu}$ pairs in the paired Fermi. 
In comparison with the ground state, for breaking $N_{ub}$ pairs, while forming $N_{bu}$ pairs, the particle numbers are given by 
$$
    N=N_G, \quad M=M_G -N_{ub} +N_{bu}, \quad 
    M_n=0, \text{ for } n=1,2,\cdots.
$$
We introduce $\delta$-functions for  representing the excited pairs  and unpaired particles and holes as
\begin{equation}
    \begin{aligned}
            &-\sum_{i=1}^{N_{ub}}\delta\left(k-\lambda_{h,i}\right)/L
            +\sum_{i=1}^{N_{ub}}\left[\delta\left(k-k_{p1,i}\right)/L
            +\delta\left(k-k_{p2,i}\right)/L\right]\text{ (for breaking $N_{ub}$ pairs) }\\
            &+\sum_{i=1}^{N_{bu}}\delta\left(k-\lambda_{p,i}\right)/L
            -\sum_{i=1}^{N_{bu}}\left[\delta\left(k-k_{h1,i}\right)/L
            +\delta\left(k-k_{h2,i}\right)/L\right]\text{ (for forming $N_{bu}$ pairs)},
    \end{aligned}
    \label{dealtabreakform}
\end{equation}
respectively, where $k_{p1,i}$, $k_{p2,i}$, $k_{h1,i}$, and $k_{h2,i}$ stand for the excited quasimomenta of unpaired particles and holes. 
As we did in Eq.(\ref{distributionintroducedelta}), substitute the $\delta$-funciotns(\ref{dealtabreakform}) into the distribution functions Eq.(\ref{eqdistribution}).  By taking similar calculation  to that of the  single particle-hole excitation, the excited energy  of multiple  pair- forming and breaking is given by  
\begin{equation}
    \begin{aligned}
        \Delta \Omega=&\int_{Q_-}^{Q_+}\rho\left(k\right) 
        \left(k^{2}-\mu_0 H\right) \mathrm{d} k
        -\int_{-Q_0}^{Q_0} \rho^{0}\left(k\right) \left(k^{2}-\mu_0 H\right) \mathrm{d} k \\
        &+2\int_{B_-}^{B_+} \sigma\left(\lambda\right) \left(\lambda^2-c^2\right) \mathrm{d}\lambda 
        -2\int_{-B_0}^{B_0}\sigma\left(\lambda\right)\left(\lambda^2-c^2\right) \mathrm{d}\lambda\\
        =&\sum_{i=1}^{N_{ub}}
        \left[\kappa^0\left(k_{p1,i}\right)+\kappa^0\left(k_{p2,i}\right)
        -\varepsilon^0\left(\lambda_{h,i}\right)\right]
        +\sum_{i=1}^{N_{bu}}
        \left[\varepsilon^0\left(\lambda_{p,i}\right)-\kappa^0\left(k_{h1,i}\right)
        -\kappa^0\left(k_{h2,i}\right)\right]\textcolor{blue}{.}
        \label{excitedenergybreakingforming}
    \end{aligned}
\end{equation}
The total  momentum for this excitation, depends on not only the quasimomenta of exited holes and particles, but also  the numbers of breaking pairs and forming pairs  $N_{ub}$ and $N_{bu}$.  
The parity of quasimomenta of pairs in such a type of excitations does not change because  it only depends on $N=N_G$ (see Eq.(\ref{eqsconfiguration})). Meanwhile the parity of quasimomenta of unpaired fermions is changed since it depends on $N+M=N_G+M_G -N_{ub} +N_{bu}$. 
Therefore, an additional phase shift  is caused  due to the change of parity, namely, 
\begin{equation}
    \begin{aligned}
        \Delta K=&
        2\pi\sum_{i=1}^{N_{ub}}\left[
        \left(\int_{0}^{k_{p1,i}}+\int_{0}^{k_{p2,i}}\right) \rho^0\left(k\right) \mathrm{d}k
        -\int_{0}^{\lambda_{h,i}} \sigma^0\left(k\right) \mathrm{d}k \right]\\&
        +2\pi\sum_{i=1}^{N_{bu}}\left[
        \int_{0}^{\lambda_{p,i}} \sigma^0\left(k\right) \mathrm{d}k
        -\left(\int_{0}^{k_{h1,i}}+\int_{0}^{k_{h2,i}}\right) \rho^0\left(k\right) \mathrm{d}k 
        \right]\pm 
        \begin{cases}
            0 &,\text {for $N_{ub}-N_{bu}$ even}\\
            \pi n_u &,\text {for $N_{ub}-N_{bu}$ odd}
        \end{cases}
        .
    \end{aligned}
    \label{excitedmomentumbreakingforming}
\end{equation}
We see that odd $\{N_{ub}-N_{bu}\}$ gives a two-fold-degenerate excited state. 
In Fig. \ref{Nub} we present excitation  spectra of breaking or forming one pair. 
Obviously, the spectra are more complicated due to the two-fold degeneracy. 

\subsection{n-strings Excitations}
\label{sectionnstring}
\label{section-strings}
\begin{figure}[t]
    \centering
    \subfloat[Dressed energies (left) and distribution functions (right) for $n=1,2,3$]{
        \includegraphics[width=0.485\textwidth]{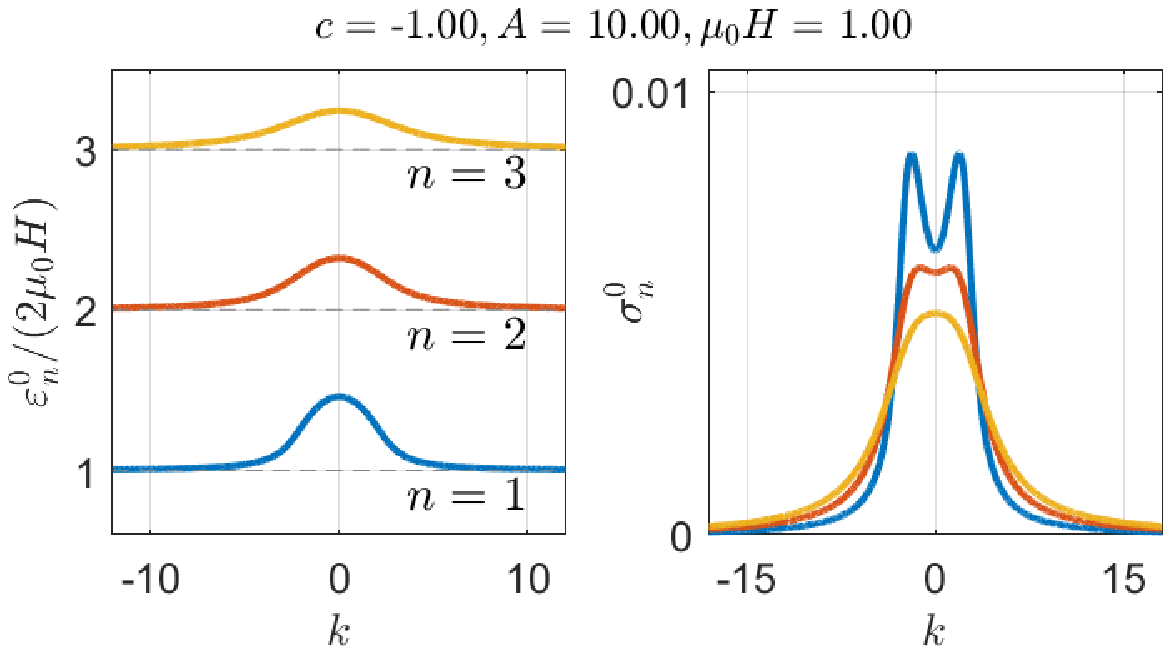} 
        \label{ndisdre}
    }
    \subfloat[Spectra of one $n$-string (left) and two length-$1$ strings (right)]{
        \includegraphics[width=0.485\textwidth]{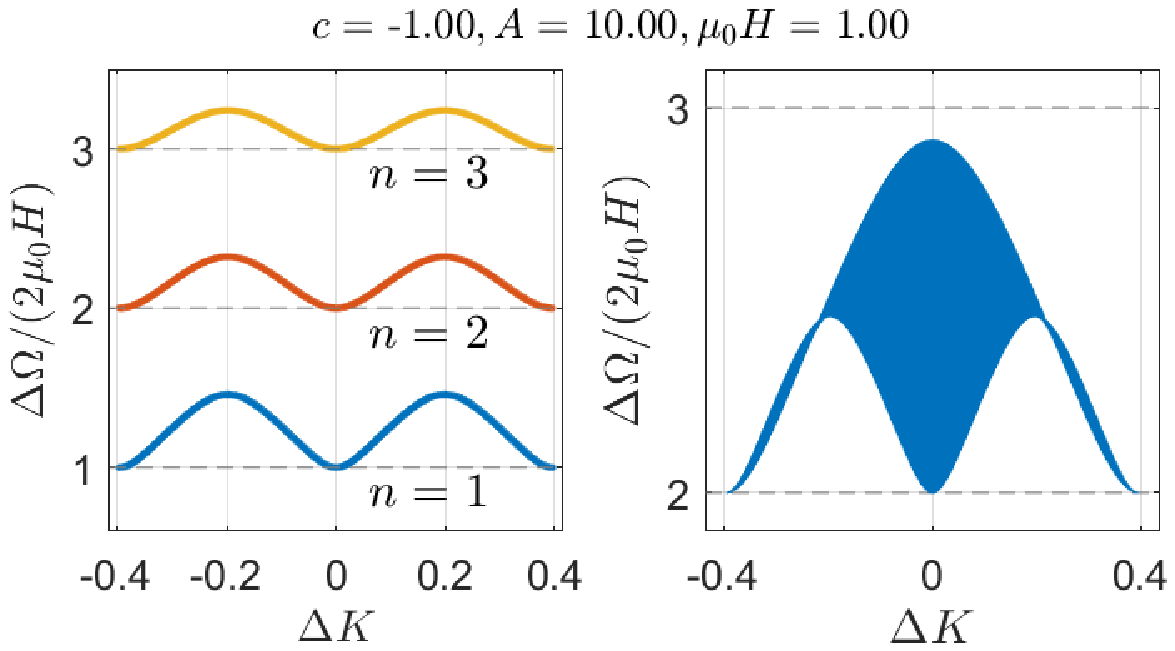} 
        \label{nstringex}
    }
    \caption[distribution functions $\sigma_n^0\left(k\right)$ 
    and dressed energies $\varepsilon_n^0\left(k\right)$ for $n=1,2,3$.]
    {\textbf{(a) Dressed energies $\varepsilon_n^0\left(k\right)$ and distribution functions $\sigma_n^0\left(k\right)$ 
    for $n=1,2,3$}. 
    The dressed energies $\varepsilon_n^0\left(k\right)$ tend to $2n\mu_0H$ for $k\rightarrow\pm\infty$. 
    For  the ground state, $\sigma_n^0$ give the distribution of holes because of $M_n=0$. 
    \textbf{(b) Excitation spectra of the spinon-bound-states }.
    For a single length-$n$ string excitation, double degeneracy spectrum is observed (left). Right figure shows  the continuum  spectra of two length-$1$ spinons.
    }
\end{figure}

Now we consider excitations by flipping   spins in the  unpaired fermions without forming pairs. 
According to Eq.(\ref{6.10}), the dressed energy and distribution function in the spin sector at the ground state are given by 
\begin{equation}
    \varepsilon_n^0\left(\lambda\right)=
    2n\mu_0H-\int_{-Q_0}^{Q_0} a_n\left(\lambda-k'\right) \kappa^0\left(k'\right) \mathrm{d}k',\quad
    \sigma_{n}^0\left(\lambda\right)=\sigma_{n}^0{ }^{h}\left(\lambda\right)=
    \int_{-Q_0}^{Q_0} a_n\left(\lambda-k'\right) \rho^0\left(k'\right) \mathrm{d}k'.
\end{equation}
For the ground state, all spin strings are unoccupied at the zero temperature. 
The dressed energies and distribution functions for $n$-strings with $n=1,2,3$ are plotted in Fig. \ref{ndisdre}.

For simplicity, consider the case of exciting one length-$l$ string by flipping  unpaired fermions, i.e., $M_l=M_{l,G}+1=1$ and $M_n=0$ for $n\neq l$.
The quantum numbers of the spin setor $J_\alpha{}^n$ are integer (half-odd integer) for $N-M_{n}$ odd (even), and satisfy Eq.(\ref{J^n}).
Therefore, for the ground state, $J_\alpha{}^n$ are confined  by the condition 
\begin{equation}
    \left|J_\alpha{}^n\right|\leqq \left(N-2 M\right)/2, 
\end{equation}
which indicates $N-2M$ holes and no occupation of all $n$-strings.
For the excited state with $M_\ell =1$, $J_\alpha{}^n$ are confined by the condition 
\begin{equation}
    \left|J_\alpha{}^n\right|\leqq 
    \begin{cases}
        \left(N-2 M-2n\right)/2 &,n<\ell,\vspace{5pt}\\
        \left(N-2 M-2\ell+1\right)/2 &,n=\ell,\vspace{5pt}\\
        \left(N-2 M-2\ell \right)/2 &,n>\ell,
    \end{cases}
\end{equation}
that  indicates $N-2M-2n$ holes for $n\leq \ell$, $N-2M-2\ell $ holes for $n>\ell $ and  one  occupation of length-$\ell$ string  for $n=\ell$.
By introduce $\delta\left(k-\lambda_{\ell}\right)/L$ representing the excited $\ell$-string into the distribution functions, we have 
\begin{equation}
    \begin{aligned}
        &\bar{\sigma}\left(k\right)=\frac{1}{\pi} 
        - \int_{-B}^{B} a_2\left(k-k'\right)\sigma\left(k'\right)\mathrm{d}k'
        - \int_{-Q}^{Q} a_1\left(k-k'\right)\rho\left(k'\right)\mathrm{d}k', \\
        &\bar{\rho}\left(k\right)=\frac{1}{2 \pi}
        - \int_{-B}^{B} a_1\left(k-k'\right)\sigma\left(k'\right)\mathrm{d}k'
        - \int_{-B_n}^{B_n} a_\ell \left(k-k'\right)\bar{\sigma}_l\left(k'\right)\mathrm{d}k'
        + \frac{1}{L}a_\ell\left(k-\lambda_{\ell }\right), \\
        &\bar{\sigma}_n\left(k\right)=
        \int_{-Q}^{Q} a_n\left(k-k'\right)\rho\left(k'\right)\mathrm{d}k',
    \end{aligned}
    \label{distributionbyintroducingdelta}
\end{equation}
where we denoted $\bar{\sigma}_n=\sigma_n+\sigma_n^h$, and we define $B_n$ by $\bar{\sigma}_n(B_n)=0$. 
The excited free energy density is thus  given by
\begin{equation}
    \begin{aligned}
        \frac{\Delta \Omega}{L}
        =&\int_{Q_-}^{Q_+}{\rho}\left(k\right) 
        \left(k^{2}-\mu_0 H\right) \mathrm{d} k 
        +2\int_{B_-}^{B_+} {\sigma}\left(k\right) \left(k^2-c^2\right) 
        \mathrm{d}k 
        +\sum_{n=1}^{\infty} 2n\mu_0H \int_{B_{n+}}^{B_{n-}} \sigma_n\left(k\right) \mathrm{d} k\\
        &-\int_{-Q_0}^{Q_0} \rho^{0}\left(k\right) \left(k^{2}-\mu_0 H\right) \mathrm{d} k
        -2\int_{-B_0}^{B_0}\sigma^0\left(k\right)\left(k^2-c^2\right) \mathrm{d}k 
        -\sum_{n=1}^{\infty} 2n\mu_0H \int_{B_{n0}}^{B_{n0}} \sigma_n^0\left(k\right) \mathrm{d} k, 
    \end{aligned}
    \label{excitedenergydensitystring}
\end{equation}
where the last term equals zero since all strings are unoccupied at the ground state. By the calculations similar to that of the excited energy of one particle-hole excitation (see Appendix \ref{Express the excited energy in dressed energies}), the excited energy of the length-$\ell$ string excitation is given by 
\begin{equation}
    \begin{aligned}
        \Delta \Omega
        =&2l\mu_0H-\int_{-Q_0}^{Q_0}a_l\left(k-\lambda_{l}\right)\kappa^{0}\left(k\right)
        \mathrm{d}k
        =\varepsilon_l^0\left(\lambda_{l}\right).
    \end{aligned}
    \label{excitedenergyonestring}
\end{equation}
In the case of exciting one length-$\ell$ string, the parity of quasimomentum  $k_j$ for the unpaired charge sector is changed due to its dependence on the quantum  numbers $M+\sum_{n=1}^\infty M_n$. 
Therefore,  apart from the contribution of the excited string, there is an additional term in the total excited momentum attributed to the momentum change of the unpaired sector, namely, 
\begin{equation}
    \Delta K
    =\frac{2\pi}{L}J_p{}^\ell \pm \pi(N-2M)/L
    =2\pi\int_{0}^{\lambda_{l}} \sigma_l^0\left(k\right) \mathrm{d}k
    \pm \pi n_u. 
\end{equation}
 From Eq. (\ref{totalmomentum}), we see that  the momentum is confined in an interval given by taking $\lambda_{\ell }\rightarrow\pm\infty$, i.e., 
\begin{equation}
    \left| \Delta K \right|
    <2\pi\int_{0}^{\infty}  \sigma_l^0\left(k\right)  \mathrm{d}k+\pi n_u
    =2\pi\int_{-Q}^{Q} \left( \int_{0}^{\infty} a_n\left(k-k'\right) \mathrm{d}k \right) \rho\left(k'\right)\mathrm{d}k'+\pi n_u=2\pi n_u.
    \label{excitedkstringmax} 
\end{equation}

In the case of  excitations of $M_n^{\mathrm{ex}}$ length-$n$ strings, i.e.  $M_n=M_n^{\mathrm{ex}}$, 
we introduce the excitation function  $\sum_{n=1}^\infty\sum_{i=1}^{M_n}\delta\left(k-\lambda_{n,i}\right)/L$ in the BA equations.
One can prove that the excited energy is the sum of those of individual $n$-string excitation, while the excited energy has an additional term due to the change of configurations in the unpaired quasimomenta
\begin{equation}
    \Delta \Omega=\sum_{n=1}^\infty\sum_{i=1}^{M_n}
    \varepsilon_n^0\left(\lambda_{n,i}\right),\quad
    \Delta K=
    2\pi\sum_{n=1}^\infty\sum_{i=1}^{M_n}\int_{0}^{\lambda_{n,i}} \sigma_n^0\left(k\right) \mathrm{d}k
    \pm
    \begin{cases}
        0 &,\text {for $\sum_{n=1}^{\infty} M_n$ even}\\
        \pm \pi n_u &,\text {for $\sum_{n=1}^{\infty} M_n$ odd}
    \end{cases}.
    \label{excitedenergymultistring}
\end{equation}
The excitation spectra of  one and two $n$-strings with $n=1,2,3$ are given in Fig. \ref{nstringex}, resepctively. 
These spectra are gapped with a magnitude of $2n\mu_0H$ over the ground state. 
Since a certain number of unpaired fermions are flipped, these excitations should be regarded as magnon excitations that show a ferromagnetic coupling in the unpaired sector. 
The string structure of these kinds  has been confirmed  in the experiment  \cite{wang2018experimental}. 

\section{Conclusion and discussion}
We have rigorously studied the excitation spectra of the Yang-Gaudin model with an attractive interaction. 
For the one particle-hole excitations in paired and unpaired  fermions, the spectra apparently manifest a novel separation of collective motions of bosonic modes, i.e., the pairs-unpaired-fermions  separation, which can be regarded as an evidence for the existence of the FFLO-state. 
We further analytically characterized this separation by calculating the curvature corrections to the linear dispersions of both paired and unpaired sections for a  small momentum and  a  strong coupling regime.
We have determined the sound velocities and effective masses in these dispersions, and compared with numerical results. 
We have showed that the free energy and the specific heat of the system can be simply expressed in a sum of inverse sound velocities of paired and unpaired Fermi seas at low temperatures,  reflecting universal thermodynamics of TLLs. 
We have also studied other types of  excitations, such as breaking or forming a pair and length-$n$ string excitations in the spin sector, in order to understand subtle novel pairing and deparing  states. 
Here the gapped length-$n$ string excitations manifest a ferromagnetic coupling in the unpaired Fermi sea,  i.e.,  magnon excitations. 

Building on our  results obtained, we expect to further study dynamical correlation functions for  the charge-charge separation theory of FFLO states.  
According to the linear response theory \cite{kubo1957statistical}, one can measure the spectra of pairing and depairing in  the system by imposing a perturbation onto the system and observing their  linear responses. 
The Bragg spectroscopy \cite{veeravalli2008bragg} is one of a common experimental tools  used to measure the many-body correlation. 
Using the Bragg spectroscopy, the recent experiment\cite{senaratne2021spin}  did by Hulet's group at Rice University  have confirmed the spin-charge separation theory of TLLs in  the 1D repulsive Fermi gas.
In the Bragg spectroscopy, $^6Li$ atoms are trapped in 1D-tubes and two beams of different orientations are imposed onto the tubes. 
Due to the disturbance of the two beams, 
initially symmetrically distributed particles distribute asymmetrically in the tube. 
The dynamic structure factor (DSF) $S(q, \omega)$, usually defined as the density-density correlation, is related to the change of the total momentum as $\Delta P \propto[S(q, \omega)-S(-q,-\omega)]$. 
Therefore, the DSF is a measurable quantity in the experiment with both repulsive and attractive Fermi gases. 
Moreover, at low-energy excitations, 
the maximum (peak) of the DSF is related to  the sound velocity, namely, the corresponding peak frequency is given by $\omega_p=v_cq$, and it is independent of the effective mass.
Accordingly, different sound velocities of the paired and unpaired fermions  can be observed by the Bragg spectroscopy, showing the subtle nature of the FFLO states in 1D.

\section*{Ackonwledgement }
J.J.L. and X.W.G. are  supported by the NSFC key grant No. 12134015, the NSFC grant  No. 11874393 and No. 12121004.
J.F.P.  thanks the X.W.G.'s team members for helpful discussion and suggestion in completing his  Master project and thank  Innovation Academy for Precision Measurement Science and Technology, Chinese Academy of Sciences for kind  hospitality. 

\newpage 

\begin{center}
\textbf{\large Appendix }
\end{center}

\setcounter{equation}{0} \setcounter{figure}{0} 
\makeatletter

\renewcommand{\thefigure}{A\arabic{figure}} \renewcommand{\thesection}{SM%
\arabic{section}} \renewcommand{\theequation}{A\arabic{equation}}

\subsection{Express excited energies in dressed energies}
\label{Express the excited energy in dressed energies}
First, introduce two useful formulas.

Formula 1: Any two groups of the analytical functions of the form
\begin{equation}
\begin{aligned}
    &\begin{cases}
        \eta\left(k\right)=\eta^{0}\left(k\right)
        -\int_{-\lambda_{0}}^{\lambda_{0}} a_{1}\left(k-\lambda\right) 
        \gamma\left(\lambda\right) \mathrm{d} \lambda \vspace{5pt}\\ 
        \gamma\left(\lambda\right)=\gamma^{0}\left(\lambda\right)
        -\int_{-k_{0}}^{k_{0}} a_{1}\left(k-\lambda\right) \eta\left(k\right) \mathrm{d} k
        -\int_{-\lambda_{0}}^{\lambda_{0}} a_{2}\left(\lambda-\lambda^{\prime}\right) 
        \gamma\left(\lambda^{\prime}\right) \mathrm{d} \lambda^{\prime}
    \end{cases} \vspace{5pt}\\
    &\begin{cases}
        f\left(k\right)=f_{0}\left(k\right)
        -\int_{-\lambda_{0}}^{\lambda_{0}} a_{1}\left(k-\lambda\right) g\left(\lambda\right) 
        \mathrm{d} \lambda \vspace{5pt} \\
        g\left(\lambda\right)=g_{0}\left(\lambda\right)
        -\int_{-k_{0}}^{k_{0}} a_{1}\left(\lambda-k\right) f\left(k\right) \mathrm{d} k
        -\int_{-\lambda_{0}}^{\lambda_{0}} a_{2}\left(\lambda-\lambda^{\prime}\right) 
        g\left(\lambda^{\prime}\right) \mathrm{d} \lambda^{\prime}
    \end{cases}
\end{aligned}
\end{equation}
satisfy
\begin{equation}
    \int_{-k_{0}}^{k_{0}} f_{0}\left(k\right) \eta\left(k\right) \mathrm{d} k
    +\int_{-\lambda_{0}}^{\lambda_{0}} g_{0}\left(\lambda\right) \gamma\left(\lambda\right) 
    \mathrm{d} \lambda
    =\int_{-k_{0}}^{k_{0}} \eta^{0}\left(k\right) f\left(k\right) \mathrm{d} k
    +\int_{-\lambda_{0}}^{\lambda_{0}} \gamma^{0}\left(\lambda\right) g\left(\lambda\right) 
    \mathrm{d} \lambda. 
\end{equation}
One can prove it by multipy $\eta\left(k\right)$ and $f\left(k\right)$.

Formula 2: For any even analytical function $f\left(k\right)$ integrated between 
$\left[k_{-}, k_{+}\right]$,where $k_{+}=k_{0}+\Delta Q_{+}, k_{-}=$ $-k_{0}+\Delta Q_{-}$
(for $\Delta Q_{\pm}$ small enough), it can be expanded with $\Delta Q_{\pm}$, i.e.,
$$
\begin{aligned}
    &\int_{k_{-}}^{k_{+}} f\left(k\right) \mathrm{d} k
    =\int_{-k_{0}+\Delta Q_{-}}^{k_{0}+\Delta Q_{+}} f\left(k\right) \mathrm{d} k\\
    =&\int_{-k_{0}}^{k_{0}} f\left(k\right) \mathrm{d} k
    +\left[f\left(k_{0}\right) \Delta Q_{+}-f\left(-k_{0}\right) \Delta Q_{-}\right]
    +\mathcal{O}\left(\left(\Delta Q\right)^{2}\right), 
\end{aligned}
$$
so that
$$
    \begin{aligned}
        &\int_{k_{-}}^{k_{+}} a_{n}\left(k-k^{\prime}\right) f\left(k^{\prime}\right) 
        \mathrm{d} k^{\prime}\\
        =&\int_{-k_{0}}^{k_{0}} a_{n}\left(k-k^{\prime}\right) 
        f\left(k^{\prime}\right) \mathrm{d} k+\left[a_{n}\left(k-k_{0}\right) \Delta Q_{+}
        -a_{n}\left(k+k_{0}\right) \Delta Q_{-}\right] f\left(k_{0}\right)
        +\mathcal{O}\left(\left(\Delta Q\right)^{2}\right)
    \end{aligned}
$$
If $k_{+}=-k_{-}=k_{0}+\Delta Q$, namely, $\Delta Q_{+}=-\Delta Q_{-}=\Delta Q$, 
the expansion can be expressed as
$$
    \int_{k-}^{k_{+}} f\left(k\right) \mathrm{d} k=
    \int_{-k_{0}-\Delta Q}^{k_{0}+\Delta Q} f\left(k\right) \mathrm{d} k
    =\int_{-k_{0}}^{k_{0}} f\left(k\right) \mathrm{d} k+2 f\left(k_{0}\right) \Delta Q
    +\mathcal{O}\left(\left(\Delta Q\right)^{2}\right)
$$
and
$$
\begin{aligned}
    &\int_{k_{-}}^{k_{+}} a_{n}\left(k-k^{\prime}\right) f\left(k^{\prime}\right) 
    \mathrm{d} k^{\prime}\\
    =&\int_{-k_{0}}^{k_{0}} a_{n}\left(k-k^{\prime}\right) f\left(k^{\prime}\right) 
    \mathrm{d} k'
    +\left[a_{n}\left(k-k_{0}\right)+a_{n}\left(k+k_{0}\right)\right] f\left(k_{0}\right) 
    \Delta Q
    +\mathcal{O}\left(\left(\Delta Q\right)^{2}\right).
\end{aligned}
$$

Introducing $-\delta(k-k_h)/L$ and $\delta(k-k_p)/L$ into the distribution functions of unpaired fermions for one-particle hole excitations, we obtain
\begin{equation}
    \begin{aligned}
        \bar{\sigma}\left(\lambda\right)=&\frac{1}{\pi} 
        - \int_{B_-}^{B_+} a_2\left(\lambda-k'\right)\sigma\left(k'\right)\mathrm{d}k'
        - \int_{Q_-}^{Q_+} a_1\left(\lambda-k'\right)\bar{\rho}\left(k'\right)\mathrm{d}k'\\
        &+\frac{1}{L}a_1\left(\lambda-k_h\right)-\frac{1}{L}a_1\left(\lambda-k_p\right) \\
        \bar{\rho}\left(k\right)=&\frac{1}{2 \pi} 
        - \int_{B_-}^{B_+} a_1\left(k-k'\right)\sigma\left(k'\right)\mathrm{d}k'.
    \end{aligned}
\end{equation}
According to Eq.(\ref{excitedenergytoberewritten}), the excited energy density is given by 
\begin{equation}
    \begin{aligned}
        \frac{\Delta \Omega}{L}
        =&\int_{Q_-}^{Q_+}\left[\bar{\rho}\left(k\right)
        -\frac{1}{L} \delta\left(k-k_{h}\right)+\frac{1}{L} \delta\left(k-k_{p}\right)\right] 
        \left(k^{2}-\mu_0 H\right) \mathrm{d} k
        -\int_{-Q_0}^{Q_0} \rho^{0}\left(k\right) \left(k^{2}-\mu_0 H\right) \mathrm{d} k \\
        &+2\int_{B_-}^{B_+} \sigma\left(\lambda\right) \left(\lambda^2-c^2\right) \mathrm{d}\lambda 
        -2\int_{-B_0}^{B_0}\sigma\left(\lambda\right)\left(\lambda^2-c^2\right) \mathrm{d}\lambda \\
        =&\int_{-Q_0}^{Q_0} \Delta \bar{\rho}\left(k\right) \left(k^{2}-\mu_0 H\right) \mathrm{d} k
        +2 \bar{\rho}\left(Q_0\right) \left(Q_0^{2}-\mu_0 H\right) \Delta Q-\frac{1}{L} k_{h}^{2}
        +\frac{1}{L} k_{p}^{2}\\
        &+2\int_{-B_0}^{B_0} \Delta \bar{\sigma}\left(\lambda\right) \left(\lambda^{2}-c^2\right) 
        \mathrm{d} \lambda
        +4 \bar{\sigma}\left(B_0\right) \left(B_0^{2}-c^2\right) \Delta B,
    \end{aligned}
\end{equation}
where $\Delta\bar{\rho}\left(k\right)=\bar{\rho}\left(k\right)-\rho^0\left(k\right)$,
$\Delta \bar{\sigma}\left(\lambda\right)
=\bar{\sigma}\left(\lambda\right)-\sigma^0\left(\lambda\right)$, 
and Formula.2 is used in the last step. Since the particle number does not change, i.e., $N-N_G=0$, we directly have
\begin{equation}
    \begin{aligned}
        0=&\int_{Q_-}^{Q_+} \rho\left(k\right) \mathrm{d} k
        -\int_{-Q_0}^{Q_0} \rho^{0}\left(k\right) \mathrm{d} k 
        +2\int_{B_-}^{B_+} \sigma\left(\lambda\right)\mathrm{d}\lambda
        -2\int_{-B_0}^{B_0} \sigma^0\left(\lambda\right)\mathrm{d}\lambda\\
        =&\int_{-Q_0}^{Q_0} \Delta\bar{\rho}\left(k\right) \mathrm{d} k
        +2 \bar{\rho}\left(Q_0\right) \Delta Q-\frac{1}{L}+\frac{1}{L}
        +2\int_{-B_0}^{B_0} \Delta \bar{\sigma}\left(\lambda\right)\mathrm{d}\lambda
        +4 \bar{\sigma}\left(B_0\right)\Delta B.
    \end{aligned}
\end{equation}
Substitute the last equation into the excited energy density, we obtain 
\begin{equation}
    \begin{aligned}
        \frac{\Delta \Omega}{L} 
        =&\int_{-Q_0}^{Q_0} \Delta \bar{\rho}\left(k\right)\left(k^{2}-A-\mu_0H\right) \mathrm{d} k
        +2\int_{-B_0}^{B_0} \Delta \bar{\sigma}\left(\lambda\right) \left(\lambda^{2}-A-c^2\right)
         \mathrm{d} \lambda \\
        &+A\left[\frac{1}{L}-\frac{1}{L}-2 \bar{\rho}\left(Q_0\right) \Delta Q
        -4 \bar{\sigma}\left(B_0\right)\Delta\lambda \right]\\
        &+2 \bar{\rho}\left(Q_0\right) \left(Q_0^{2}-\mu_0 H\right) \Delta Q
        +4 \bar{\sigma}\left(B_0\right) \left(B_0^{2}-c^2\right) \Delta B
        -\frac{1}{L} k_{h}^{2} +\frac{1}{L} k_{p}^{2} \\
        =&\int_{-Q_0}^{Q_0} \Delta \bar{\rho}\left(k\right)\left(k^{2}-A-\mu_0H\right) \mathrm{d} k
        +2\int_{-B_0}^{B_0} \Delta \bar{\sigma}\left(\lambda\right) \left(\lambda^{2}-A-c^2\right)
         \mathrm{d} \lambda \\
        &+2\left(Q_0^{2}-A-\mu_0H\right) \bar{\rho}\left(Q_0\right) \Delta Q
        +4\left(B_0^2-A-c^2\right) \bar{\sigma}\left(B_0\right) \Delta B\\
        &-\frac{1}{L}\left(k_{h}^{2}-A-\mu_0H\right)
        +\frac{1}{L}\left(k_{p}^{2}-A-\mu_0H\right).
    \end{aligned}
\end{equation}
According to Fomula.2, the change of distribution functions are given by
\begin{equation}
    \begin{aligned}
        \Delta \bar{\sigma} \left(\lambda\right)=& \bar{\sigma} \left(\lambda\right)
        -\sigma^{0}\left(\lambda\right) \\
        =& -\int_{-Q_0}^{Q_0} a_{1}\left(\lambda-k\right) \Delta \bar{\rho}\left(k\right) 
        \mathrm{d} k
        -\int_{-B_0}^{B_0} a_{2}\left(\lambda-\lambda^{\prime}\right) 
        \Delta \bar{\sigma} \left(\lambda^{\prime}\right) \mathrm{d} \lambda\\
        &-\left[a_{1}\left(\lambda-Q_0\right)+a_{1}\left(\lambda+Q_0\right)\right] 
        \bar{\rho}\left(Q_0\right) \Delta Q
        -\left[a_{2}\left(\lambda-B_0\right)+a_{2}\left(\lambda+B_0\right)
        \right] \bar{\sigma} \left(B_0\right) \Delta B\\
        &+\frac{1}{L} [a_{1}\left(\lambda-k_{h}\right)-a_{1}\left(\lambda-k_{p}\right)] \\
        \Delta \bar{\rho}\left(k\right)=& \bar{\rho}\left(k\right)-\rho^{0}\left(k\right)\\
        =& -\int_{-B_0}^{B_0} a_{1}\left(k-\lambda\right) 
        \Delta \bar{\sigma} \left(\lambda\right) 
        \mathrm{d} \lambda
        -\left[a_{1}\left(k-B_0\right)+a_{1}\left(k+B_0\right)\right] 
        \bar{\sigma} \left(B_0\right) \Delta B. \\
    \end{aligned}
\end{equation}
While according to Fomula.1, we obtain the following relation, 
\begin{equation}
    \begin{aligned}
        &2 \int_{-B_0}^{B_0} \Delta \bar{\sigma}\left(\lambda\right)\left(k^2-A-c^2\right)
        \mathrm{d}\lambda
        +\int_{-Q_0}^{Q_0} \Delta \bar{\rho}\left(k\right)\left(k^{2}-A-\mu_0 H\right) 
        \mathrm{d} k\\
        =&\int_{-B_0}^{B_0} \frac{1}{L} [a_{1}\left(\lambda-k_{h}\right)
        -a_{1}\left(\lambda-k_{p}\right)] \varepsilon^{0}\left(\lambda\right) 
        \mathrm{d} \lambda\\
        &-\int_{-B_0}^{B_0}\left[a_{1}\left(\lambda-Q_0\right)
        +a_{1}\left(\lambda+Q_0\right)\right] \varepsilon^{0}\left(\lambda\right) 
        \mathrm{d} \lambda 
        \bar{\rho}\left(Q_0\right) \Delta Q \\
        &-\int_{-B_0}^{B_0}\left[a_{2}\left(\lambda-B_0\right)
        +a_{2}\left(\lambda+B_0\right)\right] \varepsilon^{0}\left(\lambda\right) \mathrm{d} \lambda 
        \bar{\sigma}\left(B_0\right) \Delta B \\
        &-\int_{-Q_0}^{Q_0}\left[a_{1}\left(k-B_0\right)
        +a_{1}\left(k+B_0\right)\right] \kappa^{0}\left(k\right) \mathrm{d} k 
        \bar{\sigma}\left(B_0\right) \Delta B \\
    \end{aligned}
    \label{eqappendix67}
\end{equation}
Since $\varepsilon^0\left(B_0\right)=\varepsilon^0\left(-B_0\right)=0$ and 
$\kappa^0\left(Q_0\right)=\kappa^0\left(-Q_0\right)=0$, we directly obtain
\begin{equation}
    \begin{aligned}
        0=&\varepsilon^0\left(B_0\right)+\varepsilon^0\left(-B_0\right)
        =&&4\left(B_0^2-A-c^2\right)
        -\int_{-B_0}^{B_0}\left[a_{2}\left(\lambda-B_0\right)+a_{2}\left(\lambda+B_0\right)\right] \varepsilon^{0}\left(\lambda\right) \mathrm{d} \lambda \\
        &&&-\int_{-Q_0}^{Q_0}\left[a_{1}\left(k-B_0\right)
        +a_{1}\left(k+B_0\right)\right] \kappa^{0}\left(k\right) \mathrm{d} k\\
        0=&\kappa^0\left(Q_0\right)+\kappa^0\left(-Q_0\right)
        =&&2\left(Q_0^{2}-A-\mu_0H\right)
        -\int_{-B_0}^{B_0}\left[a_{1}\left(\lambda-Q_0\right)
        +a_{1}\left(\lambda+Q_0\right)\right] \varepsilon^{0}\left(\lambda\right) 
        \mathrm{d} \lambda
    \end{aligned}
    \label{relation}
\end{equation}
Therefore, the last three terms of R.H.S. of Eq.(\ref{eqappendix67}) can be substituted by the above relation (\ref{relation}), 
namely,
\begin{equation}
    \begin{aligned}
        &2 \int_{-B_0}^{B_0} \Delta \bar{\sigma}\left(\lambda\right)\left(k^2-A-c^2\right)
        \mathrm{d}\lambda
        +\int_{-Q_0}^{Q_0} \Delta \bar{\rho}\left(k\right)\left(k^{2}-A-\mu_0 H\right) 
        \mathrm{d} k\\
        =&\int_{-B_0}^{B_0} \frac{1}{L} a_{1}\left(\lambda-k_{h}\right) 
        \varepsilon^{0}\left(\lambda\right) \mathrm{d} \lambda
        -\int_{-B_0}^{B_0} \frac{1}{L} a_{1}\left(\lambda-k_{p}\right) 
        \varepsilon^{0}\left(\lambda\right) \mathrm{d} \lambda\\
        &-2\left(Q_0^{2}-A-\mu_0H\right) \bar{\rho}\left(Q_0\right) \Delta Q
        -4\left(B_0^2-A-c^2\right) \bar{\sigma}\left(B_0\right) \Delta B. \\
    \end{aligned}
\end{equation}
Substitute the last equation into the excited energy, it follows that 
\begin{equation}
    \begin{aligned}
        \Delta \Omega
        =&\int_{-B_0}^{B_0}  a_{1}\left(\lambda-k_{h}\right) 
        \varepsilon^{0}\left(\lambda\right) \mathrm{d} \lambda
        -\int_{-B_0}^{B_0}  a_{1}\left(\lambda-k_{p}\right) 
        \varepsilon^{0}\left(\lambda\right) \mathrm{d} \lambda
        -\left(k_{h}^{2}-A-\mu_0H\right)
        +\left(k_{p}^{2}-A-\mu_0H\right)\\
        =&\kappa^0\left(k_p\right)-\kappa^0\left(k_h\right).
    \end{aligned}
\end{equation}
Together with Eq.(\ref{excitedmomentumunpaired}), the dispersion of one-particle-hole excitation of the unpaired fermions is obtained.

The dispersion of paired fermions can be acquired in the same way. 
In brief, we again introduce two $\delta$-functions $-\delta\left(\lambda-\lambda_{h}\right)+\delta\left(\lambda-\lambda_{p}\right)$ as $\Delta\bar{\sigma}(\lambda)$. Accordingly, the distirbution funcitons are 
\begin{equation}
    \begin{aligned}
        \bar{\sigma}\left(\lambda\right)=&\frac{1}{\pi} 
        - \int_{B_-}^{B_+} a_2\left(\lambda-k'\right)\sigma\left(k'\right)\mathrm{d}k'
        - \int_{Q_-}^{Q_+} a_1\left(\lambda-k'\right)\bar{\rho}\left(k'\right)\mathrm{d}k'
        +\frac{1}{L}a_1\left(\lambda-\lambda_h\right)-\frac{1}{L}a_1\left(\lambda-\lambda_p\right) \\
        \bar{\rho}\left(k\right)=&\frac{1}{2 \pi} 
        - \int_{B_-}^{B_+} a_1\left(k-k'\right)\sigma\left(k'\right)\mathrm{d}k'
        +\frac{1}{L}a_1\left(\lambda-\lambda_h\right)-\frac{1}{L}a_1\left(\lambda-\lambda_p\right) .
    \end{aligned}
\end{equation}
According to Eq.(\ref{excitedenergytoberewritten}), the excited energy density is given by 
\begin{equation}
    \begin{aligned}
        \frac{\Delta \Omega}{L}
        =&\int_{-Q_0}^{Q_0} \Delta \bar{\rho}\left(k\right) \left(k^{2}-\mu_0 H\right) \mathrm{d} k
        +2 \bar{\rho}\left(Q_0\right) \left(Q_0^{2}-\mu_0 H\right) \Delta Q\\
        &+2\int_{-B_0}^{B_0} \Delta \bar{\sigma}\left(\lambda\right) \left(\lambda^{2}-c^2\right) 
        \mathrm{d} \lambda
        +4 \bar{\sigma}\left(B_0\right) \left(B_0^{2}-c^2\right) \Delta B
        -\frac{1}{L} \lambda_{h}^{2}+\frac{1}{L} \lambda_{p}^{2}.
    \end{aligned}
\end{equation}
Using the  same technique  as that for unpaired fermions, one can  essentially acquire the excited exergy given by Eq.(\ref{eqdispersionpaired}).

In general, we can have excitations of $2N_b$ particle-holes in the paired sector and $2N_u$ in the unpaired sector simultaneously. 
One can easily prove that the multiple particle-hole excitations are simply a summation of single-particle-hole excitations with $2N_u+2N_b$ $\delta$-functions introduced into the distribution functions, i.e. $\sum_{i=1}^{N_b}[-\delta\left(k-k_{h,i}\right)+\delta\left(k-k_{p,i}\right)]/L$ and $\sum_{j=1}^{N_u}[-\delta\left(k-\lambda_{h,j}\right)+\delta\left(k-\lambda_{p,j}\right)]/L$. 

For solely  breaking and forming pairs, since there is no excitation of length-$n$ strings, the dispersions are  similar to the above ones.
The difference is likely that the excited state might be degenerate in this case due to the change of particle numbers of each sector, as described by Eq.(\ref{excitedmomentumbreakingforming}).

For excitations of one  length-$l$ string, the distribution functions are given by Eq.(\ref{distributionbyintroducingdelta}). 
Since there is exactly one occupied string, $\sigma_n(k)=\delta\left(k-\lambda_{l}\right)/L$.  
Then the excited free energy density Eq.(\ref{excitedenergydensitystring}) can be rewritten as 
\begin{equation}
    \begin{aligned}
        \frac{\Delta \Omega}{L}
        =&\int_{Q_-}^{Q_+}\bar{\rho}\left(k\right) 
        \left(k^{2}-\mu_0 H\right) \mathrm{d} k
        -\int_{-Q_0}^{Q_0} \rho^{0}\left(k\right) \left(k^{2}-\mu_0 H\right) \mathrm{d} k \\
        &+2\int_{B_-}^{B_+} \bar{\sigma}\left(\lambda\right) \left(\lambda^2-c^2\right) 
        \mathrm{d}\lambda 
        -2\int_{-B_0}^{B_0}\sigma^0\left(\lambda\right)\left(\lambda^2-c^2\right) \mathrm{d}\lambda 
        +\frac{2l\mu_0H}{L} \int_{B_{l+}}^{B_{l-}} \delta\left(k-\lambda_{l}\right) \mathrm{d} k\\
        =&\int_{-Q_0}^{Q_0} \Delta \bar{\rho}\left(k\right) \left(k^{2}-A-\mu_0 H\right) \mathrm{d} k
        +2 {\rho}\left(Q_0\right) \left(Q_0^{2}-A-\mu_0 H\right) \Delta Q\\
        &+2\int_{-B_0}^{B_0} \Delta \bar{\sigma}\left(\lambda\right) \left(\lambda^{2}-A-c^2\right) 
        \mathrm{d} \lambda
        +4 \sigma\left(B_0\right) \left(B_0^{2}-A-c^2\right) \Delta B+\frac{2l\mu_0H}{L}.
    \end{aligned}
\end{equation}
Since the particle number is fixed, i.e., $N-N_G=0$, we directly obtain
\begin{equation}
    \begin{aligned}
        \frac{\Delta \Omega}{L} 
        =&\int_{-Q_0}^{Q_0} \Delta \bar{\rho}\left(k\right)\left(k^{2}-A-\mu_0H\right) \mathrm{d} k
        +2\int_{-B_0}^{B_0} \Delta \bar{\sigma}\left(\lambda\right) \left(\lambda^{2}-A-c^2\right)
         \mathrm{d} \lambda \\
        &+2\left(Q_0^{2}-A-\mu_0H\right) \rho\left(Q_0\right) \Delta Q
        +4\left(B_0^2-A-c^2\right) \sigma\left(B_0\right) \Delta B
        +\frac{2l\mu_0H}{L}.
    \end{aligned}
    \label{excitedfreeenergyinter}
\end{equation}
According to Fomula.2, the changes of distribution functions are 
\begin{equation}
    \begin{aligned}
        \Delta \bar{\sigma} \left(\lambda\right)
        =& \bar{\sigma} \left(\lambda\right)
        -\sigma^{0}\left(\lambda\right) 
        = -\int_{-Q_0}^{Q_0} a_{1}\left(\lambda-k\right) \Delta \bar{\rho}\left(k\right) 
        \mathrm{d} k
        -\left[a_{1}\left(\lambda-Q_0\right)+a_{1}\left(\lambda+Q_0\right)\right] 
        {\rho}\left(Q_0\right) \Delta Q \\
        &-\int_{-B_0}^{B_0} a_{2}\left(\lambda-\lambda^{\prime}\right) 
        \Delta \bar{\sigma} \left(\lambda^{\prime}\right) \mathrm{d} \lambda
        -\left[a_{2}\left(\lambda-B_0\right)+a_{2}\left(\lambda+B_0\right)
        \right] \sigma \left(B_0\right) \Delta B\\
        \Delta \bar{\rho}\left(k\right)
        =& \bar{\rho}\left(k\right)-\rho^{0}\left(k\right)
        = -\int_{-B_0}^{B_0} a_{1}\left(k-\lambda\right) \Delta \bar{\sigma} \left(\lambda\right) 
        \mathrm{d} \lambda
        -\left[a_{1}\left(k-B_0\right)+a_{1}\left(k+B_0\right)\right] 
        \sigma \left(B_0\right) \Delta B. \\
    \end{aligned}
\end{equation}
According to Fomula.1 and the expression of derssed energies, we have
\begin{equation}
    \begin{aligned}
        &2 \int_{-B_0}^{B_0} \Delta \bar{\sigma}\left(\lambda\right)\left(k^2-A-c^2\right)
        \mathrm{d}\lambda
        +\int_{-Q_0}^{Q_0} \Delta \bar{\rho}\left(k\right)\left(k^{2}-A-\mu_0 H\right) 
        \mathrm{d} k\\
        =&-\int_{-B_0}^{B_0}\left[a_{1}\left(\lambda-Q_0\right)
        +a_{1}\left(\lambda+Q_0\right)\right] \varepsilon^{0}\left(\lambda\right) 
        \mathrm{d} \lambda 
        {\rho}\left(Q_0\right) \Delta Q \\
        &-\int_{-B_0}^{B_0}\left[a_{2}\left(\lambda-B_0\right)
        +a_{2}\left(\lambda+B_0\right)\right] \varepsilon^{0}\left(\lambda\right)
        \mathrm{d}\lambda 
        \sigma\left(B_0\right) \Delta B \\
        &-\int_{-Q_0}^{Q_0}\left[a_{1}\left(k-B_0\right)
        +a_{1}\left(k+B_0\right)\right] \kappa^{0}\left(k\right) \mathrm{d} k 
        \sigma\left(B_0\right) \Delta B 
        -\int_{-Q_0}^{Q_0}\frac{1}{L}a_l\left(k-\lambda\right)\kappa^{0}\left(k\right) 
        \mathrm{d} k.
    \end{aligned}
\end{equation}
Since $\varepsilon^0\left(B_0\right)+\varepsilon^0\left(-B_0\right)=0$ and 
$\kappa^0\left(Q_0\right)+\kappa^0\left(-Q_0\right)=0$,
the frist three terms of R.H.S. of the last equatoin can be rewritten. Namely, we have 
\begin{equation}
    \begin{aligned}
        &2 \int_{-B_0}^{B_0} \Delta \bar{\sigma}\left(\lambda\right)\left(k^2-A-c^2\right)
        \mathrm{d}\lambda
        +\int_{-Q_0}^{Q_0} \Delta \bar{\rho}\left(k\right)\left(k^{2}-A-\mu_0 H\right) 
        \mathrm{d} k\\
        =&-\int_{-Q_0}^{Q_0}\frac{1}{L}a_l\left(k-\lambda\right)\kappa^{0}\left(k\right)
        \mathrm{d}k. 
        -2\left(Q_0^{2}-A-\mu_0H\right) {\rho}\left(Q_0\right) \Delta Q
        -4\left(B_0^2-A-c^2\right) \sigma\left(B_0\right) \Delta B\\
        \end{aligned}
\end{equation}
Substituting the last equation into Eq.(\ref{excitedfreeenergyinter}) we obtain the excited energy Eq.(\ref{excitedenergyonestring}) expressed in the length-$\ell$-string dressed energy.

\subsection{Sound Velocity and Effective Mass from Bethe Ansatz}
\label{appendixvcm}
In Section \ref{sectionvm}, we derived sound velocities and effective masses of paired 
and unpaired fermions from the zero-temperature TBA equations. 
Here we  apply the method of Ref.\cite{he2020emergence} to derive the  results from the BA equations in detail.

Since $M_n=0$ at the ground state, the logarithm of the BA equations of the two charge sectors are
\begin{equation}
    \begin{aligned}
        &2 \Lambda_{\alpha} L=2 \pi J_{\alpha}+
        \sum_{j=1}^{N-2 M} 
        \theta\left(\frac{\Lambda_{\alpha}-k_{j}}{|c|}\right)+
        \sum_{\beta=1}^{M} 
        \theta\left(\frac{\Lambda_{\alpha}
        -\Lambda_{\beta}}{2|c|}\right), &
        & \alpha=1,2, \cdots, M \\
        &k_{j} L=2 \pi I_{j}+
        \sum_{\alpha=1}^{M} 
        \theta\left(\frac{k_{j}-\Lambda_{\alpha}}{|c|}\right), &
        & j=1, \cdots, N-2 M \\
    \end{aligned}
    \label{1st2BA}
\end{equation}
For the ground state, at the thermodynamic limit, 
we can expand both equations with coupling strength up to order $\sim \mathcal{O}(1 / c)$, 
the first of which gives
$$
\begin{aligned}
    \Lambda_\alpha 
    \approx&\frac{\pi}{L} J_\alpha
    +\frac{1}{2L} \sum_{j=1}^{N-2M} 2 
    \left[\arctan\left(\frac{-k_j}{\left|c\right|}\right)
    +\frac{1}{1+\left(k_j/ c\right)^{2}} \frac{ \Lambda_\alpha}{\left|c\right|}
    \right] \\
    &+\frac{1}{2L} \sum_{\beta=1}^{M} 2 
    \left[\arctan\left(\frac{-\Lambda_\beta}{2\left|c\right|}\right)
    +\frac{1}{1+\left(\Lambda_\beta/c\right)^{2}/4} \frac{ \Lambda_\alpha}{2\left|c\right|}
    \right]. 
\end{aligned}
$$
If $k_j$ and $\Lambda_\beta$ are both symmetrically distributed
to the zero-point, the $\arctan$-terms drop out in the sum with respect to $j$ and $\beta$. Namely, we have
\begin{equation}
\begin{aligned}
    \Lambda_\alpha =\frac{\pi}{L} J_\alpha
    &+\frac{1}{L} \sum_{j=1}^{N-2M} 
    \frac{1}{1+\left(k_j/c\right)^{2}} \frac{\Lambda_\alpha}{\left|c\right|}
    +\frac{1}{L} \sum_{\beta=1}^{M} 
    \frac{1}{1+\left(\Lambda_\beta/c\right)^{2}/4} \frac{\Lambda_\alpha}{2\left|c\right|} \\
    =\frac{\pi}{L} J_\alpha
    &+\int_{-Q}^{Q} \frac{1}{1+\left(k_j/c\right)^{2}} 
    \rho\left(k\right)\mathrm{d}k \frac{\Lambda_\alpha}{\left|c\right|} 
    +\int_{-B}^{B} \frac{1}{1+\left(\lambda/c\right)^{2}/4} 
    \sigma\left(\lambda\right)\mathrm{d}\lambda \frac{\Lambda_\alpha}{2\left|c\right|} \\
    =\frac{\pi}{L} J_\alpha
    &+\int_{-Q}^{Q} \frac{1}{1+\left(k_j/c\right)^{2}} 
    \left[\frac{1}{2 \pi}-a_1\left(k\right)n_b\right]\mathrm{d}k 
    \frac{\Lambda_\alpha}{\left|c\right|} \\
    &+\int_{-B}^{B} \frac{1}{1+\left(\lambda/c\right)^{2}/4} 
    \left[\frac{1}{\pi}-a_2\left(\lambda\right)n_b - a_1\left(\lambda\right)n_u\right]
    \mathrm{d}\lambda \frac{\Lambda_\alpha}{2\left|c\right|} \\
    \approx \frac{\pi}{L} J_\alpha
    &+\left[\frac{\left|c\right|}{\pi}\arctan\left(\frac{Q}{\left|c\right|}\right)
    +\frac{2\left|c\right|}{\pi}\arctan\left(\frac{B}{2\left|c\right|}\right)\right]
    \frac{\Lambda_\alpha}{\left|c\right|} \\
    \approx \frac{\pi}{L} J_\alpha&+\frac{Q+B}{\pi \left|c\right|}\Lambda_\alpha ;\\
\end{aligned}
\label{lambdaalpha}
\end{equation}
if $k_j$ or $\Lambda_\beta$ is asymmetrically distributed to the zero-point, 
there is an additional but negligible term   
\begin{equation}
    \pm\frac{1}{L}\arctan(\frac{\pi n_u}{\left|c\right|})
    \approx \pm\frac{\pi n_u}{L\left|c\right|}\ll \frac{\pi}{L}J_\alpha \quad\text{or}\quad
    \pm\frac{1}{L}\arctan(\frac{\pi n_b}{\left|c\right|})
    \approx \pm\frac{\pi n_b}{L\left|c\right|}\ll \frac{\pi}{L}J_\alpha,
    \label{lambdaalphaasymmtry}
\end{equation}
respectively.
According to Eq.(\ref{lambdaalpha}) and (\ref{lambdaalphaasymmtry}), the approximation of $\Lambda_\alpha$ is independent of the parity of $k_j$.
By substituting Eq.(\ref{BQinnbnu}) into Eq.(\ref{lambdaalpha}), we have
\begin{equation}
    \Lambda_\alpha
    =\left(1-\frac{Q+B}{\pi \left|c\right|}\right)^{-1} 
    \frac{\pi}{L} J_\alpha
    \approx\left(1+\frac{Q+B}{\pi \left|c\right|}\right) 
    \frac{\pi}{L} J_\alpha
    \approx\left(1+\frac{1}{2\gamma_b}+\frac{1}{\gamma_u}\right)\frac{\pi}{L} J_\alpha,
\end{equation}
where the first term of RHS of the last equation $\pi J_\alpha/L$ equals the quasimomentum of free paired fermions, while the remaining terms emerge from the interactions among fermions.

The total energy and momentum of the paired fermions are given by
\begin{equation}
    E_{b}=2\sum_{\alpha=1}^{M} \Lambda_\alpha^{2}
    =2\left(1+\tau_b\right)^{2}\left(\frac{\pi}{L}\right)^{2} \sum_{\alpha=1}^{M} J_\alpha^{2},\quad
    K_{b}=2\sum_{\alpha=1}^{M} \Lambda_\alpha
    =2\left(1+\tau_b\right) \frac{\pi}{L} \sum_{\alpha=1}^{M} J_\alpha.
\end{equation}
where $\tau_b=\frac{1}{2\gamma_b}+\frac{1}{\gamma_u}$. 
For one-particle-hole excitation near the Fermi surface, the quantum number of the excited particle and hole are
\begin{equation}
    \left|J_p\right|>\frac{M-1}{2},   \quad 
    \left|J_h\right|=\frac{M-1}{2}-\beta<\frac{M-1}{2}, \text{ }
    \beta\in\mathbb{Z}^+\text{ and }\beta\ll M,
\end{equation}
where $\beta$ is chosen to be much smaller than $M$ so that the excitation is taken near the Fermi surface.
Therefore, the excited momentum of free paired fermions is
\begin{equation}
    \Delta K_{b,0}=2\frac{\pi}{L}\left(J_p-J_h\right);
\end{equation}
whereas the excited energy of attractive paired fermions is
\begin{equation}
    \begin{aligned}
        \Delta E_{b} 
        &=2\left(1+\tau_b\right)^{2}\left(\frac{\pi}{L}\right)^{2}\left(J_p^2-J_h^2\right) 
        =2\left(1+\tau_b\right)^{2}\left(\frac{\pi}{L}\right)^{2} 
        \frac{L \Delta K_{b,0}}{2\pi}\left[\left(M-1-2\beta\right)+\frac{L 
        \Delta K_{b,0}}{2\pi}\right] \\
        &=\left(1+\tau_b\right)^{2} \pi n_{b} \Delta K_{b,0}
        +\frac{\left(1+\tau_b\right)^{2}}{2}\left(\Delta K_{b,0}\right)^{2} 
        +\mathcal{O}\left(\frac{1}{L}\right).
        \label{paireddispersion}
    \end{aligned}
\end{equation}
We would like to express this dispersion in free-fermion approximation, with the mass of free fermion $m$ (set to be $1/2$ previously) replaced by the effective mass $m^*$ due to the interactions. Namely, 
\begin{equation}
    \Delta E_{b} 
    =2\left[v_{b} \frac{\Delta K_{b,0}}{2}
    +\frac{1}{2 m_b^*}\left(\frac{\Delta K_{b,0}}{2}\right)^{2}\right],
    \label{pairedfreefermionapproximation}
\end{equation}
By comparison of Eq,(\ref{paireddispersion}) and (\ref{pairedfreefermionapproximation}), the sound velocities and effective masses are given by
\begin{equation}
    \begin{aligned}
        &v_{b}=\pi n_{b}\left(1+\frac{1}{2\gamma_b}+\frac{1}{\gamma_u}\right)^{2} 
        \approx \pi n_{b}\left(1+\frac{1}{\gamma_b}+\frac{2}{\gamma_u}\right),\\
        &m^{*}_b=m\left(1+\frac{1}{2\gamma_b}+\frac{1}{\gamma_u}\right)^{-2} 
        \approx m\left(1-\frac{1}{\gamma_b}-\frac{2}{\gamma_u}\right),
    \end{aligned}
\end{equation}
which is the same as the sound velocity and effective mass of paired fermions in 
Eq.(\ref{vbmb}).

Similarly, we can calculate these quantities of unpaired sectors. 
By expanding the second equation of Eq.(\ref{1st2BA}), we have
\begin{equation}
    \begin{aligned}
        k_{j} \approx \frac{2 \pi}{L} I_{j}+\frac{4B}{\pi \left|c\right|}k_j
        \approx \left(1+\frac{4B}{\pi \left|c\right|}\right) \frac{2\pi}{L} I_j
        \approx \left(1+\frac{2}{\gamma_b}\right) \frac{2\pi}{L} I_j.
    \end{aligned}
\end{equation}
The dispersion of unpaired fermions can be expressed as 
\begin{equation}
    \begin{aligned}
        \Delta E_{u} 
        &=\left(1-\tau_u\right)^{2} 2\pi n_{u} \Delta K_{u,0}
        +\left(1-\tau_u\right)^{2}\left(\Delta K_{u,0}\right)^{2} 
        +\mathcal{O}\left(\frac{1}{L}\right)\\
        &=v_{u} \Delta K_{u,0}
        +\frac{1}{2 m^{*}_u}\left(\Delta K_{u,0}\right)^{2}.
    \end{aligned}
\end{equation}
Therefore, the sound velocity and effective mass of unpaired fermions are
\begin{equation}
    \begin{aligned}
        &v_{u}=2\pi n_{u}\left(1+\frac{2}{\gamma_b}\right)^{2} 
        \approx 2\pi n_{u}\left(1+\frac{4}{\gamma_b}\right),\\
        &m^{*}_u=m\left(1+\frac{2}{\gamma_b}\right)^{-2} 
        \approx m\left(1-\frac{4}{\gamma_b}\right), 
    \end{aligned}
\end{equation}
which is the same as the sound velocity and effective mass of unpaired fermions in 
Eqs. (\ref{vumu}).

\subsection{Free Energy at Low Temperature}
\label{Free Energy at Low-Temperature}
To distinguish quantities at zero-temperature and low-temperature, we introduce the subscript "0" indicating zero-temperature. 
Accordingly, dressed energies at zero-temperature $\varepsilon_0$ and $\kappa_0$ given by Eq.(\ref{eqdressed1}); 
while dressed energies at low-temperature can be expanded with respect to zero-temperature as $\varepsilon=\varepsilon_0+\beta$ and $\kappa=\kappa_0+\gamma$. Namely, we have 
\begin{equation}
    \begin{aligned}
        \varepsilon(k)=&2(k^2-A-c^2)
        +T\int a_2(k-k')\ln{(1+e^{-\frac{\varepsilon(k')}{T}})}\mathrm{d}k'
        +T\int a_1(k-k')\ln{(1+e^{-\frac{\kappa(k')}{T}})}\mathrm{d}k'\\
        \approx& 2(k^2-A-c^2)
        +T\int a_2(k-k')\ln{(1+e^{-\frac{\left|\varepsilon(k')\right|}{T}})}\mathrm{d}k'
        -\int_{-B}^{B}a_2(k-k')\varepsilon(k')\mathrm{d}k'\\
        &+T\int a_1(k-k')\ln{(1+e^{-\frac{\left|\kappa(k')\right|}{T}})}\mathrm{d}k'
        -\int_{-Q}^{Q}a_1(k-k')\kappa(k')\mathrm{d}k'\\
        =&\varepsilon_0(k)
        +T\int a_2(k-k')\ln{(1+e^{-\frac{\left|\varepsilon(k')\right|}{T}})}\mathrm{d}k'
        -\int_{-B}^{B}a_2(k-k')\beta(k')\mathrm{d}k'\\
        &+T\int a_1(k-k')\ln{(1+e^{-\frac{\left|\kappa(k')\right|}{T}})}\mathrm{d}k'
        -\int_{-Q}^{Q}a_1(k-k')\gamma(k')\mathrm{d}k',\\
        \kappa(k)=&k^2-A-\mu_0H
        +T\int a_1(k-k')\ln{(1+e^{-\frac{\varepsilon(k')}{T}})}\mathrm{d}k'\\
        \approx& k^2-A-\mu_0H
        +T\int a_1(k-k')\ln{(1+e^{-\frac{\left|\varepsilon(k')\right|}{T}})}\mathrm{d}k'
        -\int_{-B}^{B}a_1(k-k')\varepsilon(k')\mathrm{d}k'\\
        =&\kappa_0(k)
        +T\int a_1(k-k')\ln{(1+e^{-\frac{\left|\varepsilon(k')\right|}{T}})}\mathrm{d}k'
        -\int_{-B}^{B}a_1(k-k')\beta(k')\mathrm{d}k',
    \end{aligned}
\end{equation}
where we omitted the integral intervals $(-\infty,\infty)$ for convenience of notation. 
Since for low-temperature, the integrals on the R.H.S. of the last two equations mostly attribute to the values of  integrands near the Fermi momenta $B$ and $Q$, we replace $\varepsilon(k)$ and $\kappa(k)$ with their linear expansions $(k-B)u_b$ and $(k-Q)u_u$, respectively, where $u_b$ and $u_u$ denote $\varepsilon'(B)$ and $\kappa'(Q)$, respectively. Namely, we have 
\begin{equation}
    \begin{aligned}
        \varepsilon(k)=\varepsilon_0+\beta
        =&\varepsilon_0(k)
        +\frac{\pi^2 T^2}{6u_b}\left[a_2(k-B)+a_2(k+B)\right]
        -\int_{-B}^{B}a_2(k-k')\beta(k')\mathrm{d}k'\\
        &+\frac{\pi^2 T^2}{6u_u}\left[a_1(k-Q)+a_1(k+Q)\right]
        -\int_{-Q}^{Q}a_1(k-k')\gamma(k')\mathrm{d}k',\\
        \kappa(k)=\kappa_0+\gamma
        =&\kappa_0(k)
        +\frac{\pi^2 T^2}{6u_b}\left[a_1(k-B)+a_1(k+B)\right]
        -\int_{-B}^{B}a_1(k-k')\beta(k')\mathrm{d}k'.
    \end{aligned}
\end{equation}
We further denote $\beta_0(k)=(\pi^2 T^2)/(6u_b)\left[a_2(k-B)+a_2(k+B)\right]+(\pi^2 T^2)/(6u_u)\left[a_1(k-Q)+\right.$ $\left.a_1(k+Q)\right]$ and $\gamma_0(k)=(\pi^2 T^2)/(6u_b)\left[a_1(k-B)+a_1(k+B)\right]$, so that the last equations can be simplified as 
\begin{equation}
    \begin{aligned}
        \varepsilon(k)=\varepsilon_0+\beta=&\varepsilon_0+\beta_0
        -\int_{-B}^{B}a_2(k-k')\beta(k')\mathrm{d}k'
        -\int_{-Q}^{Q}a_1(k-k')\gamma(k')\mathrm{d}k',\\
        \kappa(k)=\kappa_0+\gamma=&\kappa_0+\gamma_0
        -\int_{-B}^{B}a_1(k-k')\beta(k')\mathrm{d}k'.
    \end{aligned}
    \label{dressedlowtempexpension}
\end{equation}

According to Eq.(\ref{equationofstate}), the free energy density (i.e., pressure) at zero-temperature is given by 
\begin{equation}
    \begin{aligned}
        \Omega_0/L=p_0=-\frac{1}{\pi}\int_{-B}^{B}\varepsilon_0(k)\mathrm{d}k
        -\frac{1}{2\pi}\int_{-Q}^{Q}\kappa_0(k)\mathrm{d}k;
    \end{aligned}
\end{equation}
while the free energy density at low temperature can be expanded with respect to zero-temperature by 
\begin{equation}
    \begin{aligned}
        &\Omega/L=p\\=&
        -\frac{1}{\pi}\int_{-B}^{B}\varepsilon(k)\mathrm{d}k
        -\frac{1}{2\pi}\int_{-Q}^{Q}\kappa(k)\mathrm{d}k
        +T \int \ln{(1+e^{-\frac{\left|\varepsilon(k')\right|}{T}})} \frac{\mathrm{d}k}{\pi}
        +T \int \ln{(1+e^{-\frac{\left|\kappa(k')\right|}{T}})} \frac{\mathrm{d}k}{2 \pi}\\
        =&p_0-\frac{1}{\pi}\int_{-B}^{B}\beta(k)\mathrm{d}k
        -\frac{1}{2\pi}\int_{-Q}^{Q}\gamma(k)\mathrm{d}k
        +\frac{\pi T^2}{3u_b}+\frac{\pi T^2}{6u_u}.
    \end{aligned}
    \label{appendixpressureinter}
\end{equation}
Below we will prove that the last four terms on the R.H.S can be expressed in sound velocities $v_b$ and $v_u$. 

First, we calculate $\int_{-Q}^{Q} \gamma(k)\rho(k)\mathrm{d}k$. 
According to expressions of distribution functions given by Eq.(\ref{eqdistribution}) and $\beta(k)$ and $\gamma(k)$ given by Eq.(\ref{dressedlowtempexpension}), we have 
\begin{equation}
    \begin{aligned}
        &\frac{1}{2\pi}\int_{-Q}^{Q}\gamma(k)\mathrm{d}k
        -\int_{-Q}^{Q}\int_{-B}^{B}a_1(k-k')\sigma(k')\gamma(k)\mathrm{d}k'\mathrm{d}k\\
        =&\int_{-Q}^{Q}\rho(k)\gamma_0(k)\mathrm{d}k
        +\int_{-Q}^{Q}\int_{-B}^{B}\rho(k)a_1(k'-k)\beta(k')\mathrm{d}k'\mathrm{d}k\\
        =&\int_{-Q}^{Q}\rho(k)\gamma_0(k)\mathrm{d}k
        +\int_{-B}^{B}\left[\sigma(k')-\frac{1}{\pi}+\int_{-B}^{B}a_2(k-k')\sigma(k')\mathrm{d}k\right]\beta(k')\mathrm{d}k'\\
        =&\int_{-Q}^{Q}\rho(k)\gamma_0(k)\mathrm{d}k
        +\left(\int_{-B}^{B}\sigma(k')\beta_0(k')\mathrm{d}k'
        -\int_{-B}^{B}\int_{-B}^{B}\sigma(k')a_2(k'-k)\beta(k)\mathrm{d}k\mathrm{d}k'\right.\\
        &\left.-\int_{-B}^{B}\int_{-Q}^{Q}\sigma(k')a_1(k'-k)\gamma(k)\mathrm{d}k\mathrm{d}k'\right)
        -\frac{1}{\pi}\int_{-B}^{B}\beta(k')\mathrm{d}k'
        +\int_{-B}^{B}\int_{-B}^{B}a_2(k-k')\sigma(k')\beta(k')\mathrm{d}k\mathrm{d}k',
    \end{aligned}
\end{equation}
where the last term on the L.H.S should cancel with the fourth term on the R.H.S, and the third and last terms on the R.H.S should cancel with each other. By rearranging the last equation and substituting $\beta_0$ and $\gamma_0$ into it, we have 
\begin{equation}
    \begin{aligned}
        &\frac{1}{\pi}\int_{-B}^{B}\beta(k)\mathrm{d}k
        +\frac{1}{2\pi}\int_{-Q}^{Q}\gamma(k)\mathrm{d}k
        =\int_{-B}^{B}\sigma(k)\beta_0(k)\mathrm{d}k
        +\int_{-Q}^{Q}\rho(k)\gamma_0(k)\mathrm{d}k\\
        =&\frac{\pi^2 T^2}{6u_b}\int_{-B}^{B}\sigma(k)\left[a_2(k-B)+a_2(k+B)\right]\mathrm{d}k
        +\frac{\pi^2 T^2}{6u_u}\int_{-B}^{B}\sigma(k)\left[a_1(k-Q)+a_1(k+Q)\right]\mathrm{d}k\\
        &+\frac{\pi^2 T^2}{6u_b}\int_{-Q}^{Q}\rho(k)\left[a_1(k-B)+a_1(k+B)\right]\mathrm{d}k\\
        =&\frac{\pi^2 T^2}{6u_b}\left[\frac{2}{\pi}-2\sigma(B)\right]
        +\frac{\pi^2 T^2}{6u_u}\left[\frac{1}{\pi}-2\rho(Q)\right], 
    \end{aligned}
    \label{appendixsubstituteintopressure}
\end{equation}
where the last step is because that since $a_n(k)$ is even, we have $2\sigma(B)=2/\pi-\int_{-B}^{B}\left[a_2(k-B)+\right.$ $\left.a_2(k+B)\right]\sigma(k)\mathrm{d}k-\int_{-Q}^{Q}\left[a_1(k-B)+a_1(k+B)\right]\rho(k)\mathrm{d}k$ and $2\rho(Q)=1/\pi-\int_{-B}^{B}\left[a_1(k-Q)+\right.$ $\left.a_1(k+Q)\right]\sigma(k)\mathrm{d}k$. 
By substituting Eq.(\ref{appendixsubstituteintopressure}) into Eq.(\ref{appendixpressureinter}), the free energy density can be written as 
\begin{equation}
    \begin{aligned}
        \Omega/L-\Omega_0/L=p-p_0
        =&-\frac{\pi T^2}{3u_b}+\frac{\pi^2 T^2\sigma(B)}{3u_b}
        -\frac{\pi T^2}{6u_u}+\frac{\pi^2 T^2\rho(Q)}{3u_u}
        +\frac{\pi T^2}{3u_b}+\frac{\pi T^2}{6u_u}\\
        =&\frac{\pi T^2}{6}\left(\frac{1}{v_b}+\frac{1}{v_u}\right),
    \end{aligned}
    \label{appendixpressure}
\end{equation}
where we used the relation between sound velocities and distribution functions and dressed energies given by Eqs.(\ref{appendixvcmintermpaired}) and (\ref{appendixvcmintermunpaired}), which is the same as Eq.(\ref{eqpressureinvelocity}).

\end{document}